\documentclass[prb,floatfix,showpacs,tightenlines,preprint, groupedaddress,superscriptaddress]{revtex4}
\usepackage{amsmath}
\usepackage{pslatex}
\usepackage{graphics,epsfig}
\usepackage{graphicx}
\usepackage{subfigure}
\newcommand{\smallfrac}[2]{\mbox{$\frac{#1}{#2}$}}
\newcommand{\half}{\smallfrac{1}{2} }

\newcommand{\expect}[1]{\mbox{$\langle #1 \rangle$}}
\newcommand{\fuller}{C$_{60}$ }
\newcommand{\ea}{{\em et al.} }
\renewcommand{\hbar}{{\mathchoice%
{{\raisebox{-0.33ex}{$\displaystyle\mathchar'26$}\mkern-7.2muh}}
{{\raisebox{-0.33ex}{$\displaystyle\mathchar'26$}\mkern-7.2muh}}
{{\raisebox{-0.06ex}{$\scriptstyle\mathchar'26$}\mkern-7.2muh}}
{{\raisebox{-0.02ex}{$\scriptscriptstyle\mathchar'26$}\mkern-7.2muh}}}}
\usepackage[a4paper,margin=2.7cm]{geometry}
\begin{document}

\title{Quantum Noise in the Electromechanical Shuttle}
\author{D.~Wahyu Utami}
\email{wahyu@physics.uq.edu.au} \affiliation{School of Physical
Sciences, The University of Queensland, QLD 4072, Australia}
\author{Hsi-Sheng Goan}
\email{goan@phys.ntu.edu.tw} \affiliation{Department of Physics,
National Taiwan University, Taipei 106, Taiwan, ROC}
\author{C.~A.~Holmes}
\affiliation{Department of Mathematics, School of Physical Sciences,
The University of Queensland, QLD 4072, Australia}
\author{G.~J.~Milburn}
\affiliation{Center for Quantum Computer Technology and Department
of Physics, School of Physical Sciences, The University of
Queensland, QLD 4072, Australia}

\begin{abstract}
We consider a type of Quantum Electro-Mechanical System, known as
the shuttle system, first proposed by Gorelik \ea, [Phys. Rev.
Lett., {\bf 80}, 4526, (1998)]. We use a quantum master equation
treatment and compare the semi-classical solution to a full quantum
simulation to reveal the dynamics,  followed by a discussion of the
current noise of the system. The transition between tunnelling and
shuttling regime can be measured directly in the spectrum of the
noise.
\end{abstract}
\pacs{72.70.+m,73.23.-b,73.63.Kv,62.25.+g,61.46.+w,42.50.Lc}

\maketitle

\section{Introduction}

Nanofabrication techniques, combined with single electronics, have
recently enabled position measurements on an electromechanical
oscillator to approach the Heisenberg
limit\cite{knobel-cleland,lahaye,ekinci}.  In this paper we present
a master equation treatment of a version of a quantum
electromechanical system (QEMS), the charge shuttle, first proposed
by Gorelik \cite{gorelik}. In the original  proposal a metallic
grain is surrounded by elastic soft organic molecules and placed
between two electrodes. This forms a Single Electron Transistor
(SET) with a movable island. The coupling between the vibration of
the island and the tunnelling onto the SET island dramatically
alters the transport properties of the SET.  The tunnelling
amplitudes between the reservoirs and the island are an exponential
function of the separation between island and the reservoirs.  If
the island is oscillating with a non negligible amplitude, this
separation is a function of the displacement of the island from
equilibrium and thus the tunneling current is modulated by the
motion of the island. When there is a non-zero charge on the island
the applied electric field accelerates the island. As the electron
number on the island is a stochastic quantity, the resulting applied
force is itself stochastic, but constant for a given electron
occupancy of the island.   Assuming the restoring force on the
island can be approximated as harmonic,  we have a picture of a
system moving on multiple quadratic potential surfaces, with
differing equilibrium displacements, connected by conditional
Poisson processes corresponding to tunneling of electrons on and off
the island. The shuttle thus provides a fascinating example of a
quantum stochastic system in which electron transport and
vibrational motion are strongly coupled.

In this paper we idealise the island to a single quantum dot with
only one quasi-bound electronic state. This corresponds to an
extreme Coulomb blockade regime in which the energy required for
double occupancy is not bound. This minimal model captures the
essential quantum stochastic dynamics of the shuttle system.  The
quantum dot jumps between two quadratic potential surfaces,
displaced from each other, corresponding to no electron on the
island and one electron on the island. As noted by previous authors,
the system exhibits rich dynamics including a fixed point to limit
cycle bifurcation in which the average electron occupation number on
the island exhibits a periodic  square wave dependence. In this
paper we give a quantum master equation treatment of this quantum
stochastic dynamical system, with particular attention to the
shuttling and the current noise spectrum. We use the Quantum Optics
Toolbox\cite{qotoolbox} to compare and contrast the well known
semiclassical predictions to the full quantum dynamics.  In
particular, we  compare the picture of ensemble averaged dynamics of
various moments with a `quantum trajectory'\cite{molmer} simulation
of moments. A quantum trajectory is a concept taken from quantum
optics to describe the conditional dynamics of the system
conditioned on a particular history of  stochastic events. Such
conditional dynamics provide insight into the effect of quantum
noise on the  the semiclassical prediction of regular electron
shuttling on the limit cycle.

Various versions of a charge shuttle system have been experimentally
investigated. A review of the theoretical and experimental
achievements in shuttle transport can also be found in the work of
Shekhter {\em et al.}\cite{shekhter}. When a voltage bias is applied
between the electrodes, a current quantisation resulting from
electron interactions with the vibrational levels for different
voltage bias was found. By using \fuller embedded between two gold
electrodes, Park {\em et al.} \cite{park} have demonstrated that
indeed there is current quantisation for various bias voltage which
results in a stair-like feature within the current-voltage curve.
Although because of its high frequency (around Terra Hertz) and low
amplitude oscillation, the molecule hardly shuttles between the
electrodes in this setup, this experiment has provided key evidence
of the involvement of vibrational levels in changing the properties
of the current. This quantized conductance also was observed in
several other experiments \cite{zhitenev,erbe}. Zhitenev {\em et
al.} \cite{zhitenev} utilize metal single electron transistor
attached on the tip of quartz rods as scanning probe while the
experiment by Erbe {\em et al.} \cite{erbe}, combines a
nanomechanical resonator with an electron island to produce a QEMS
system. The experimental setup used by Erbe is similar to the one
proposed by Gorelik \cite{gorelik}. Huang {\em et al.} also reported
the operation of a GHz mechanical oscillator\cite{huang}.

Several attempts to explain the behaviour of the system have been
offered both from classical and quantum point of view.  The current
quantisation and its low frequency noise was investigated via a
classical approach by Isacsson \cite{isacsson_noise}. The
current-voltage relation in the shuttle system exists within two
regimes. The first regime is when the electron tunnels straight into
the dot from the source and off to the drain, without much
involvement of the island movement. This is called the tunnel
regime. The \fuller system lies within this tunnel regime. The other
regime is when the island oscillates to accommodate the current
flow, which we call the shuttle regime. However, measurement of
average current alone cannot provide enough information to
distinguish whether the system is in the shuttle regime or tunnel
regime. It was shown that a calculation of the noise is needed in
addition. Therefore the noise signature was first obtained by
finding the Fano factor at zero frequency \cite{novotny_shot}.
Recently Flindt et al.\cite{flindt} have calculated the current
noise spectrum using a method  different form that used in this
paper. We compare the two methods in section VI.

Another interesting property of the system is the existence of a
dynamical instability with limit-cycle behavior which was found in a
similar setup using a single metallic grain placed on a cantilever
between two electrodes \cite{isacsson}. This forms a three-terminal
contact shuttle system. Classical analysis of the system points to
the fact that this instability in the system leads to deterministic
chaos. The semiclassical dynamics of the simpler case of the
isolated island, the subject of this paper, was thoroughly
investigated by Donarini et al. \cite{donarini}.

One of the early attempt to investigate the system within the
quantum limit is given by Aji {\em et al.} \cite{aji} where
electronic-vibrational coupling is investigated both in elastic and
inelastic electron transport by looking at the current-voltage
relationship and conductance. Other properties of the transport
within the shuttle system such as negative differential conductance
have also been found \cite{mccarthy} although the derivation  only
considers terms linear in the position of the island. Various
conditions, such as when the electron tunnelling length is much
greater than the amplitude of the zero point oscillations of the
central island, have been investigated by
Fedorets\cite{fedorets_shuttle}. Using phase space methods in terms
of Wigner function Novotny {\em et al.} \cite{novotny,jauho}
identify crossover from tunnelling to shuttling regime.

Another variation of the shuttle is offered by Armour and
MacKinnon \cite{armourandmackinon}. In this model the steady state current across a chain of three
quantum dots system (one dot connected to each leads and one dot as
vibrating island) was analysed by looking at the eigenspectrum.
Numerical simulation here considers 25 phonon levels, within the
large bias limit.

In a recent thesis of Donarini \cite{donarini}, the single dot
quantum shuttle and the three dot shuttle system was investigated
using Generalized Master Equation approach using Wigner distribution
functions. The current and Fano factor at zero frequency is also
investigated.

\section{The Model}

The system consists of a quantum dot 'island' moving between two
electrodes, the source and the drain.  This is analogous to a
quantum dot SET in which the island of the SET is allowed to
oscillate and thus modulate the tunnel conductance between itself
and the reservoirs. However unlike a SET we do not include a separate charging gate for the island. When a voltage bias is applied between the two electrodes, the electron from the source can tunnel onto the island
and as the island moves closer to the drain the electron can tunnel
off, thus producing a current. Here we assume that only one
electronic level is available within the island, a condition of
strong coulomb blockade.

\begin{figure}[htbp]
\centering
\includegraphics[width=8cm]{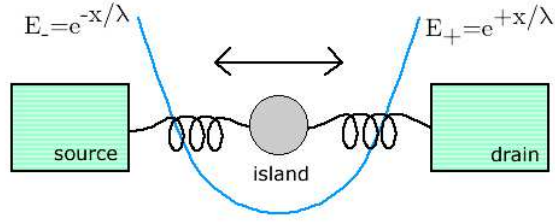}
\caption{Schematic representation of shuttling between a source and
a drain through a quantum dot.} \label{fig:shuttlesystem}
\end{figure}


The electronic single quasi-bound state on the dot is described by
Fermi  annihilation and creation operators $c,c^\dagger$, which
satisfy the anti commutation relation $cc^\dagger +c^\dagger c=1$.
While the vibrational degree of freedom is described by a
displacement operator $\hat{x}$ which can be written in terms of
annihilation and creation operators $a$ and $a^\dagger$, with the
commutation relation $aa^\dagger -a^\dagger a=1$.
\begin{equation}
\hat{x}=\sqrt{\frac{\hbar}{2m\nu}}(a+a^\dagger).
\end{equation}

The Hamiltonian of the system is given by:
\begin{eqnarray}
\lefteqn{H = \hbar \omega_I c^{\dagger}c  +  U_c \hat{n}^2} \\
& & + \hspace{0.1cm} \hbar \nu a^{\dagger}a \label{e:osc}\\
& & + \hspace{0.1cm} \hbar {\omega_sk a_k^{\dagger}a_k +  \hbar
\omega_dk b_k^{\dagger}b_k} \label{e:dns}\\
& & - \hspace{0.1cm}eE\ \hat{x}\ c^\dagger c \label{e:coupling} \\
& & + \hspace{0.1cm} \sum_k (T_{sk} E_{-}(\hat{x}) a_{k}c^{\dagger } + {\rm h.c}) + \sum_k (T_{dk}E_{+}(\hat{x}) b_{k}c^{\dagger } + {\rm h.c}) \label{e:dnscoupling} \\
& & + \hspace{0.1cm} \sum_p g (a^\dagger d_p + a d_p^\dagger) +
\sum_p \hbar \omega_p d_p^\dagger d_p, \label{e:damp}
\end{eqnarray}
where $E$ is the electric field seen by an electron on the dot.

The first term of the Hamiltonian describes the energy of a
single-electron quasi-bound state of the island. For the purpose
of our simulation, we will scale other energies in terms of this
island energy and thus conveniently set $\hbar \omega_I = 1$.  The
Coulomb charge energy, $U_c$ is the energy that is required to add
an electron when there is already one electron occupying the
island ($\hat{n} = c^\dagger c$). This energy is assumed to be
large enough so that no more than one electron occupies the island
at any time. This is the Coulomb blockade regime. In this regime
it is better to regard the island as a single quantum dot rather
than a metal island  and we will refer to it as such in the
remainder of this paper. The free Hamiltonian for the oscillator
is described in term (\ref{e:osc}) where $\nu$ is the frequency of
the mechanical oscillation of the quantum dot. The electrostatic
energy of electrons in the source ($s$) and drain ($d$) reservoirs
is written as term (\ref{e:dns}). With $a_k, a_k^\dagger$ and
$b_k, b_k^\dagger$ the annihilation and creation operator for the
electron in the source and drain respectively. Term
(\ref{e:coupling}) describes the electrostatic coupling between
the oscillator and charge while term (\ref{e:dnscoupling})
represents the source-island tunnel coupling and the drain-island
tunnel coupling. In the shuttle system, the island of the SET is
designed to move between the source and the drain terminal with an
amplitude or fluctuation comparable to the distance of the island
to the lead. Thus we introduce the term
\begin{eqnarray}
E_{\pm}(\hat{x}) & = & e^{\pm\hat{x}/\lambda} \\
            & = & e^{\pm\eta(a+a^\dagger)}
\label{exp-jump}
\end{eqnarray}
with
\begin{equation}
\eta=\left (\frac{\hbar}{2m\nu}\right )^{1/2}\frac{1}{\lambda}
\end{equation}
to account for the change in the tunnelling rate to the left and
the right lead as the position of the shuttle varies.

The last term, (\ref{e:damp}), describes the coupling between the
oscillator and the thermo-mechanical bath responsible for damping
and thermal noise in the mechanical system in the rotating wave
approximation\cite{gardiner-zoller}. We include it in order to bound
the motion under certain bias conditions.

We now obtain a closed evolution for the system of quantum dot plus
oscillator by tracing out over the degrees of freedom in the leads.
A Markov master equation for the island-oscillator system can then
be derived in the Born and Markov approximation using standard
techniques\cite{wahyu}. If we assume the vibrational frequency of
the oscillator is slow compared to bath relaxation time scales, we
arrive at:
\begin{eqnarray}
\dot{\rho}&=& -i \nu [a^\dagger a, \rho] \nonumber
\\
&+& i \chi [(a+a^\dagger) c^\dagger c, \rho] \nonumber \\
&+& \gamma _{L}\bigl( f(\hbar \omega_I - \mu_L)
\mathcal{D}[c^\dagger E_{-}(\hat{x})] \rho + (1-f(\hbar \omega_I -
\mu_L))
\mathcal{D}[c E_-(\hat{x})] \rho \bigr) \nonumber \\
&+& \gamma _{R} \bigl( f(\hbar \omega_I - \mu_R)
\mathcal{D}[c^\dagger E_+(\hat{x})] \rho + (1-f(\hbar \omega_I - \mu_R))
\mathcal{D}[c E_+(\hat{x})] \rho \bigr) \nonumber \\
&+& \kappa(\bar{n}_p+1) \mathcal{D} [a] \rho + \kappa \bar{n}_p
\mathcal{D} [a^\dagger] \rho, \label{e:Imastereqn}
\end{eqnarray}
with $\chi=eE\eta\lambda$ and $\bar{n}_p$ is the mean phonon number for the vibrational damping reservoir. We also
have defined
\begin{equation}
{\cal D}[A]\rho=A\rho A^\dagger -\frac{1}{2}(A^\dagger A\rho+\rho
A^\dagger A),
\end{equation}
where $f$ is the Fermi function $f(\epsilon)=1/(e^{\epsilon/T_{\rm
el}}+1)$. This Fermi function has an implicit dependence on the
temperature, $T_{\rm el}$, of the electronic system and the bias
conditions between the source and the drain. The terms
$\gamma_L,\gamma_R$ describe the rates of electron tunnelling form
the source to the dot and dot to drain respectively. We have
implicitly ignored co-tunnelling and higher order scattering events,
so this equation applies under weak bias and weak tunnelling
conditions. The final two terms proportional to $\kappa$ describe
the damping of the oscillator, where $\bar{n}_p= 1/(e^{\hbar
\nu/k_BT}-1)$  and $T$ are respectively the mean excitation and the
effective temperature of a thermal bath responsible for this damping
process.  Thermal mechanical fluctuations in the metal contacts of
the source and drain cause fluctuations in position of the center of
the trapping potential confining the island, that is to say small,
fluctuating linear forces act on the island.  For a harmonic trap,
this appears to the oscillator as a thermal bath. However such a
mechanism is expected to be very weak. This fact, together with the
very large frequency of the oscillator, justifies our use of the
quantum optical master equation (as opposed to the Brownian motion
master equation)  to describe this source of dissipation
\cite{gardiner-zoller}.

In order to discuss the phenomenology of this system we first
consider a special case. Under appropriate bias conditions and
very low temperature, the quasi bound state on the island is well
below the Fermi level in the source and well above the Fermi level
in the drain. The master equation then takes the `` zero
temperature'' form
\begin{eqnarray}
\dot{\rho}&=& -i \nu [a^\dagger a, \rho] + i \chi [(a+a^\dagger) c^\dagger c, \rho] \nonumber \\
&+& \gamma _{L} \mathcal{D}[c^\dagger E_-(\hat{x})] \rho + \gamma _{R}\mathcal{D}[c E_+(\hat{x})] \rho \bigr) \nonumber \\
&+& \kappa(\bar{n}_p+1) \mathcal{D} [a] \rho + \kappa \bar{n}_p
\mathcal{D} [a^\dagger] \rho. \label{e:Imastereqn2}
\end{eqnarray}

The terms proportional to $\gamma_L$ and $\gamma_R$ describe two
conditional Poisson processes, $dN_L(t),dN_R(t)$,  in which an
electron tunnels on to or off the island. The average rate of these
processes is given by \cite{goan_continuous, goan_dynamics,
goan_montecarlo}
\begin{eqnarray*}
\mathcal{E}(dN_L(t)) & =  & \gamma_L{\rm Tr}[E_-(\hat{x})c^\dagger \rho c E_-(\hat{x})]dt, \label{left_jump_ensemble}\\
\mathcal{E}(dN_R(t)) & =  & \gamma_R{\rm Tr}[E_+(\hat{x})c\rho
c^\dagger E_+(\hat{x})]dt.\label{right_jump_ensemble}
\end{eqnarray*}
where $\mathcal{E}$ refers to a classical stochastic average. Using
the cyclic property of trace and the definition in
Eq.(\ref{exp-jump}) we see that
\begin{eqnarray}
\mathcal{E}(dN_L(t)) & =  & \gamma_L\langle e^{-2\hat{x}/\lambda}cc^\dagger\rangle dt,\label{left_jump}\\
\mathcal{E}(dN_R(t)) & =  & \gamma_R\langle
e^{2\hat{x}/\lambda}c^\dagger c\rangle dt.\label{right_jump}
\end{eqnarray}
It is now possible to see that the current through the dot will depend
on the position of the oscillator.
Under appropriate operating conditions (discussed below) we can use
this dependance to configure the device as a position sensor or weak
force detector.
For a symmetric case where the tunnel-junction capacitances are
almost the same, $C_L \approx C_R$ (neglecting the position
dependence of the capacitances), the Ramo-Shockley theorem indicates
that the average current in the circuit can be given by
\begin{equation}
I(t)=\mathcal{E}(i(t))=\frac{e}{2}\left [\mathcal{E}\left
(\frac{dN_L(t)}{dt}\right ) +\mathcal{E}\left
(\frac{dN_R(t)}{dt}\right )\right ]. \label{e:averageI}
\end{equation}
If $\eta \ll 1$ and $\gamma_L = \gamma_R = \gamma$ we may write this
as
\begin{eqnarray}
I(t) & \approx & e\gamma/2+\frac{e\gamma}{\lambda}\langle\hat{x}(c^\dagger c-cc^\dagger)\rangle+\frac{e\gamma}{\lambda^2}\langle \hat{x}^2\rangle \\
& = &
e\gamma/2+\frac{e\gamma}{\lambda}(\langle\hat{x}\rangle_1-\langle\hat{x}\rangle_0)+\frac{e\gamma}{\lambda^2}\langle
\hat{x}^2\rangle ,
\end{eqnarray}
where
\begin{equation}
\langle\hat{x}\rangle_k={\rm Tr}_{osc}[\hat{x}\langle
k|\rho|k\rangle]
\end{equation}
with $k=0,1$ the occupation number states for the dot, and $osc$
indicates a trace with respect to the oscillator Hilbert space
alone. It is apparent that $\langle\hat{x}\rangle_k$ refers to the
average position of the oscillator conditioned on a particular
occupation of the dot. Clearly the average current through the
system depends on the position of the oscillating dot. However the
dependence on the first moment of position may be very weak. If
the tunnel rates through the dot are much larger than all other
time scales we expect that the occupation of the dot will reach an
equilibrium value of \half quickly. In this case the term linear
in position will be very small, leaving only a quadratic
dependence. However if it can be arranged that
$\gamma_L\neq\gamma_R$, there will be a direct dependance of the
current on the oscillator position. To clarify this situation we
first look at a semiclassical description of the dynamics.

\section{Semiclassical dynamics}
The master equation Eq.(\ref{e:Imastereqn})enables us to calculate
the coupled dynamics of the vibrational and electronic degrees of
freedom. The equations of motion for the occupation number on the
dot and the average phonon number are
\begin{eqnarray}
 \frac{d\langle c^\dagger c \rangle}{dt} &=& \gamma_L[ f_L \langle c c^\dagger e^{-2\hat{x}/\lambda} \rangle
 - (1-f_L) \langle c^\dagger c e^{-2\hat{x}/\lambda} \rangle]
\nonumber\\
& &\mbox{}+ \gamma_R[f_R( \langle c c^\dagger e^{2\hat{x}/\lambda}
\rangle - (1-f_R)\langle c^\dagger c e^{2\hat{x}/\lambda} \rangle],
\label{elect-num}\\
 \frac{d\langle a^\dagger a \rangle}{dt} &=& \gamma_L\eta^2
 [f_L  \langle c c^\dagger e^{-2\hat{x}/\lambda} \rangle + (1-f_L)\langle c^\dagger c e^{-2\hat{x}/\lambda} \rangle] \nonumber \\
 & &\mbox{}+ \gamma_R\eta^2[f_R \langle c c^\dagger e^{2\hat{x}/\lambda} \rangle
 + (1-f_R)\langle c^\dagger c e^{-2\hat{x}/\lambda}  \rangle]\nonumber \\
& &\mbox{} - i \chi \langle (a-a^\dagger) c^\dagger c \rangle+
\kappa \bar{n} - \kappa \langle a^\dagger a\rangle ,
\label{phon-num}
\end{eqnarray}
where the Fermi factors are defined by $f_\alpha  = f(\omega_I-\mu_\alpha)$ with $\alpha=L,R$ and $\mu_\alpha$ is the chemical potential in the source ($\alpha=L$) and drain ($\alpha=R$) and $\hbar\omega_I$ is the energy of the quasi bound state on the dot.
The equation of motion for the average amplitude is relatively simple:
\begin{eqnarray}
 \frac{d\langle a \rangle}{dt} &=& -i \nu \langle a \rangle - \half \kappa
 \langle a \rangle + i \chi \langle c^\dagger c \rangle
\end{eqnarray}
which is the equation of motion for a damped oscillator with time
dependent driving. Unfortunately these first order number moments
are coupled into higher order moments generating a hierarchy of
coupled equations. A semiclassical approximation to the dynamics
may be defined by factorising moments for electronic and
vibrational degrees of freedom. This discards quantum correlations
and thus is certainly not the appropriate way to describe a
quantum limited measurement. However it does enable us to see the
essential features of the dynamical character of this problem. We
will return to the full quantum problem in the next section.

We begin the semi-classical approach by factoring moments of
oscillator and electronic coordinates, for example of \expect{c
c^\dagger E_-^2} into \expect{c c^\dagger} \expect{E_-^2}, to obtain
\begin{eqnarray}
\frac{d \langle c^\dagger c \rangle}{dt} &=& \gamma_L (\langle
E_-^2 \rangle f_L - \langle c^\dagger c
\rangle \langle E_-^2 \rangle) \nonumber \\
&&+ \gamma_R (\langle E_+^2 \rangle f_R - \langle c^\dagger c
\rangle \langle E_+^2 \rangle).
\end{eqnarray}
Using the definitions,
\begin{equation}
\hat{x} = \eta \lambda (a+a^\dagger)\ \ \ \ \ \ \ \ \hat{p} = -i
\frac{\hbar}{2\eta \lambda}(a-a^\dagger)\nonumber
\end{equation}
we can write the semiclassical equations in terms of  position
$x=\langle \hat{x}\rangle$, momentum $p=\langle\hat{p}\rangle$ and
electron number $n=\langle c^\dagger c\rangle$,
\begin{eqnarray}
\frac{d n }{dt} &=& \gamma_L (e^{-2x/\lambda} f_L - ne^{-2x/\lambda})+ \gamma_R (e^{2x/\lambda} f_R - n e^{2x/\lambda}) \\
\frac{dx }{dt} &=& \frac{ p }{m} -\frac{\kappa   }{2}x\\
\frac{d p }{dt} &=& -m \nu^2 x -\frac{\kappa  }{2}p+ \chi \sqrt{2m
\nu \hbar }\ n
\end{eqnarray}
where we have made the further factorisation $\langle
E_\pm^2(\hat{x})\rangle=e^{\pm2x/\lambda}$. These results agrees
with the previous classical equations obtained by Isacsson
\cite{isacsson}, in the case of zero gate voltage on the island.
We will carefully consider the regime of validity of these
semiclassical equations in section V. For now we note that
factorising vibrational and electronic degrees of freedom ignores
any entanglement between these systems, while factorising the
exponential assumes the oscillator is very well localised in
position.

In the zero temperature limit and appropriate bias we have that $f_L=1,\ f_R=0$. The semiclassical equations of motion then take the form
\begin{eqnarray}
\frac{d n }{dt} &=& \gamma_L (1 - n)e^{-4\eta X}- \gamma_R  n e^{4\eta X} \label{e:shuttle-dyn1} \\
\frac{d\alpha}{dt} & = & -i\nu\alpha-\frac{\kappa}{2}\alpha+i\chi
n\label{e:shuttle-dyn2}
\end{eqnarray}
with
$$
\alpha=\langle a\rangle=\langle \hat{x}\rangle/(2\lambda\eta)+i\langle\hat{p}\rangle\lambda\eta/\hbar\equiv X+iY\ \ .
$$
The system of equations, Eq.(\ref{e:shuttle-dyn1},
\ref{e:shuttle-dyn2}) has a fixed point, which undergoes a hopf
bifurcation.

To see this we begin by scaling the parameters by $\nu$ and $\eta$;
$\frac{\gamma }{\nu} \rightarrow \gamma$, $\frac{\kappa }{\nu}
\rightarrow \kappa$ and $\frac{ \eta \chi }{\nu} \rightarrow \chi$
and $ \nu \rightarrow 1$ by scaling time $\tau = \nu t$  and
redefining $X$ and $Y$ by letting $\alpha = \eta ( X + i Y)$. Then
\begin{eqnarray}
\frac{d n}{d \tau} = \gamma_L ( 1-n)e^{-4 X} - \gamma_R n e^{4 X} \\
\frac{d \alpha}{d \tau} = -i \alpha - \frac{\kappa}{2} \alpha + i
\chi n .
\end{eqnarray}

The fixed point is given implicitly by
\begin{eqnarray}
n_*&=&\frac{\gamma_L e^{-4 X_*}}{\gamma_L e^{-4 X_*}+\gamma_R e^{4
X_*}}
= \frac{1}{1 + \frac{\gamma_R}{\gamma_L} e^{8 X_*}}, \label{e:fpn}\\
X_* &=& \frac{\chi }{1+ (\frac{\kappa}{2})^2}n_*  \label{e:fpx}\\
Y_* &=& \frac{\chi \frac{\kappa}{2} }{1+ (\frac{\kappa}{2})^2}n_*
\label{e:fpy}
\end{eqnarray}
from which we can see that it must satisfy,
\begin{equation}
\chi =  X_* (1+ (\frac{\kappa}{2})^2) (1 + \frac{\gamma_R}{\gamma_L}
e^{8 X_*}). \label{e:fixedX}
\end{equation}

At the hopf bifurcation the fixed point looses stability and a
limitcycle is created. To see this, first obtain the linearized
matrix about the stationary point.
$$ DF =  \left(\begin{array}{cccrrr}
    -A_*&-\frac{8 \gamma_L \gamma_R}{A_*}& 0\\ 0&-\frac{\kappa}{2}&1
    \\\chi& -1 &-\frac{\kappa}{2}
     \end{array}\right)
     $$
where $$ A_* = \gamma_L e^{-4 X_*} + \gamma_R e^{4 X_*}.$$

The stability of the fixed point is determined by the eigenvalues of
this matrix. If one or more of the eigenvalues have positive real
part  the fixed point is unstable. For complex eigenvalues the
transition between stable and unstable occurs when the eigenvalues
are pure imaginary. Here this is when
$$ \chi = \chi_h = \frac{A_* \kappa ( A_* (A_*+\kappa) +1+
  (\frac{\kappa}{2})^2)} {8 \gamma_L \gamma_R}.$$
At $\chi = \chi_h$ the eigenvalues are $ -(A_*+ \kappa), \pm i \mu $
where $\mu = \sqrt{A_* \kappa +1+ (\frac{\kappa}{2})^2}$ and the
fixed point has a one dimensional stable manifold and a two
dimensional center manifold. For $\chi < \chi_h$ the fixed point is
stable and for $\chi > \chi_h$ it is unstable. This suggests a hopf
bifurcation, however it is necessary to work out the stability
coefficient to determine if it is subcritical, creating an unstable
limitcycle or supercritical, creating a stable limitcycle. This
involves some algebra. First transform the system in the vicinity of
the fixed point to normal form via the matrix of eigenvectors $P$.
$$ P = \left(\begin{array}{cccrrr}
    8\gamma_L \gamma_R&0&8\gamma_L \gamma_R\\ A_* \kappa &- A_* \mu&- A_*^2
    \\-A_* \kappa (A_* + \frac{\kappa}{2}) &-A_* (A_* + \frac{\kappa}{2})&
    -A_*(\frac{A_* \kappa}{2} +1 +(\frac{\kappa}{2})^2)
         \end{array}\right)
          $$

Then in normal form coordinates $ \bf{u}$ $ = P^{-1}
(n-n_*,\,X-X_*,\,Y-Y_*)^T$ the system becomes
\begin{eqnarray}
 \frac{d \bf{u}}{d
\tau} = \left(\begin{array}{cccrrr} -(A_*+ \kappa)&0&0\\
0&0&-i\mu\\0&i\mu&0
 \end{array}\right) \mbox{\bf{u}} + \mbox{\bf{g}} Nl(\mbox{\bf{u}}) +
 8\gamma_L \gamma_R \mbox{\bf{f}} (\chi -\chi_h)
  (u_1+u_3),
  \end{eqnarray}
 where  $\bf{g}$ and $\bf{f}$ are column vectors, whose entries are
 $ g_i=P_{i1}^{-1},\, f_i=P_{i3}^{-1}.$ $Nl(\bf{u})$ is a scalar nonlinear function of $u_i$
obtained by perturbation. To cubic order in $u_i$
$$ Nl( {\bf u}) =  -4 n' X' \sqrt{A_*^2 - 4 \gamma_L \gamma_R}
- 8 n' {X'}^2 A_* - \frac{64
\gamma_L \gamma_R {X'}^3}{3 A_*}
$$
where
$$
(n', X', Y')^T=P \bf{u}^T.
$$
Now the limitcycle bifurcates into the center manifold which is
tangent to the $u_1=0$ plane. So if $u_1= h( u_2,\,u_3) $ is  the
equation of the center manifold through $(0,\,0,\,0)$ at $\chi =
\chi_h$, then $h(0,\,0)=0$
and $\frac{\partial h}{\partial u_i}(0,\,0) =0$. This means that a
Taylor series approximation to the center manifold will have no
constant or linear term and so the first nonzero terms are of
quadratic order in $u_i$ and
$$ h( u_2,\,u_3) = a_{20} u_2^2 + a_{11}u_2 u_3 + a_{02}u_3^2 +
\mbox{higher order terms},
$$
for some $a_{20},\, a_{11}$ and $a_{02}.$
Now differentiating $u_1= h( u_2,\,u_3) $ gives;
$$ \frac{d u_1}{d \tau} =  \frac{\partial h}{\partial u_2}\frac{d u_2}{d \tau}
+\frac{\partial h}{\partial u_3}\frac{d u_3}{d \tau}.$$ On the
center manifold $\frac{d u_i}{d \tau}(  h( u_2,\,u_3),\, u_2,\,u_3)$
are functions of $u_2$ and $u_3$ only, so this equation can be used
to calculate the coefficients $a_{i,j}$ in the Taylor series
approximation to
 $ h( u_2,\,u_3)$ recursively, by equating coefficients of like powers
of $u_2$ and $u_3$. Once  $ h( u_2,\,u_3)$ is found this can be fed
back into the equations of motion for  $u_2$ and $u_3$ to obtain the
approximate equations of motion on the center manifold. Finally the
stability coefficient for a two dimensional system in normal form
\cite{glendinning} is
  $ \displaystyle{a = \frac{1}{16}(f_{xxx} + g_{xxy}+ f_{xyy} + g_{yyy})+
   \frac{1}{16 \omega}(f_{xy}(f_{xx}+f_{yy}) -
 g_{xy}(g_{xx}+g_{yy})-f_{xx}g_{xx} + f_{yy}g_{yy})}$ evaluated at
 $(0,\,0),$
where here $\omega = \mu = \sqrt{A_* \kappa +1+
(\frac{\kappa}{2})^2}$. The subscripts indicate partial derivatives
of function $f$ or $g$ with respect to the variables $x$ and $y$.
     For instance $f_{xx}, f_{xxx}$ is is a short hand for 2nd and 3rd derivative of function
     $f$ with respect to $x$.
     Here the stability coefficient must be calculated numerically because the position of
     fixed point is only known implicitly via Eq.\ (\ref{e:fixedX}).

\begin{figure}[h!]
\centering
\includegraphics[width=8cm]{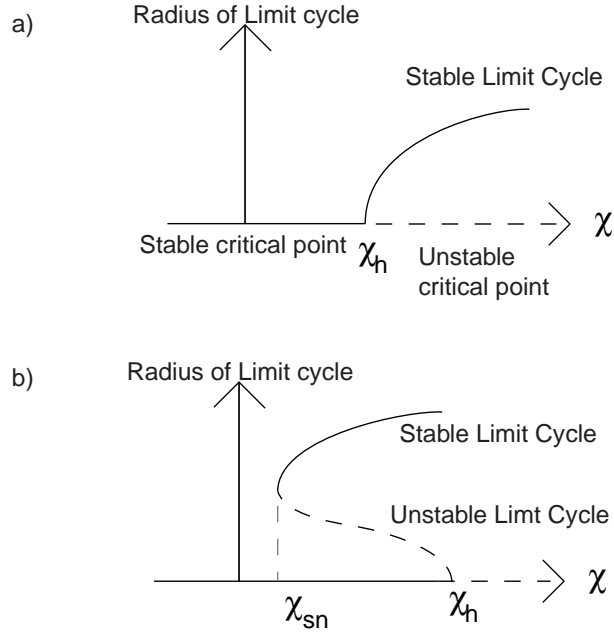}
\caption{Illustration of two possible type of Hopf bifurcation in
the shuttle system with varying coupling $\chi$, a) supercritical
and b) subcritical and saddle node bifurcation of the limit cycles.
} \label{fig:subandsuper}
\end{figure}

\begin{figure}[h!]
\centering
\includegraphics[width=10cm]{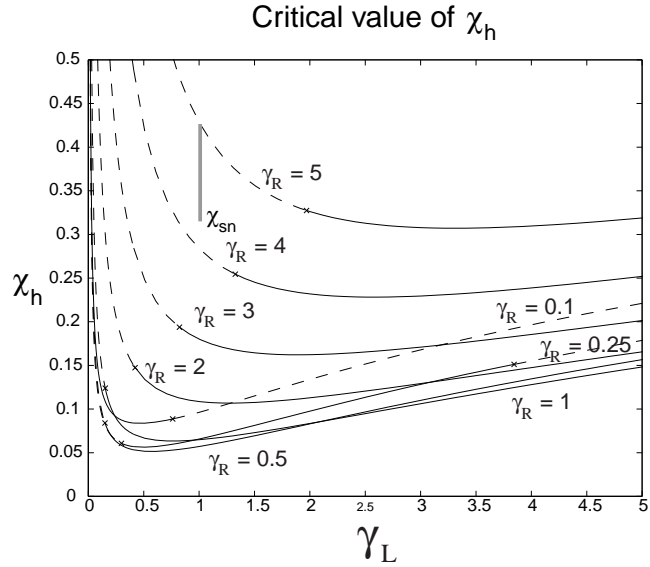}
\caption{Plot of $\chi_h$ for various fixed values of $\gamma_R$ as
a function of $\gamma_L$. The line is solid where the hopf
bifurcation is supercritical and dashed where it is a subcritical
hopf bifurcation. } \label{fig:chih}
\end{figure}

Figure \ref{fig:chih} plots $\chi_h$ for various fixed values of
$\gamma_R$ as a function of $\gamma_L$. The line is solid where the
stability coefficient is negative, implying a supercritical hopf
bifurcation and dashed where it is positive, implying a subcritical
hopf bifurcation.

At a supercritical bifurcation a stable limit cycle bifurcates from
the fixed point, existing for $\chi > \chi_h$. At a subcritical an
unstable limitcycle bifurcates from the fixed point, existing for
$\chi < \chi_h$.  Continuity of solutions as the parameter
$\gamma_L$ is changed suggests that the stable limitcycle existing
for  $\chi > \chi_h$ above the solid critical line also exists above
the dashed line. Numerical evidence shows this to be the case and
that it continues to exist well below the dashed line, eventually
being annihilated in a saddle-node bifurcation with the unstable
limit cycle created in the subcritical hopf bifurcation at the
dashed line. A schematic diagram of the two bifurcations are shown
in figure \ref{fig:subandsuper}. For $\gamma_R =5$ and $\gamma_L=
1.01$ the hopf bifurcation occurs at $ \chi_h=0. 42650636$ and the
saddle node bifurcation  at $ \chi_{sn} = 0.315. $ A glance at
Fig.\\ref{fig:chih}, where a vertical grey line indicates the range
of $ \chi$ for $\gamma_R =5$ and $\gamma_L= 1.01$ for which there
are two limit cycles shows that there is a significant parameter
region, where two limit cycles coexist.

In general  for fixed $\gamma_R$ the stability coefficient is
positive for small and very large $\gamma_L$ and negative in
between. This means that if $\gamma_L$ and  $\gamma_R$ are  about 1,
say, a stable limit cycle bifurcates and is present for $\chi
>\chi_h$. But if $({\gamma_L}/{\gamma_R})$ is much less than 1, a
more complicated situation may arise for  $\chi <\chi_h$, where an
unstable limit cycle exists close to the critical point surrounded
by a stable limit cycle.

We then solved numerically the full system of equations,
Eqs.(\ref{e:shuttle-dyn1}) and (\ref{e:shuttle-dyn2}), for various
values of the parameters.  In the shuttling regime the electron
number on the dot $n(t)$ exhibits a square wave dependence as a
single electron is carried from source to drain, where it tunnels
onto the drain and the dot returns empty to the source to repeat the
cycle. This is shown as the thin line in Fig.\ref{fig:qoshuttle}(a).
 The effect of shuttling generally occurs when the maximum
 displacement of the island is quite
large, and where the strength of the tunneling depends strongly on
the position of the island ($\lambda$ small). During shuttling, the
electron number on the dot is constant. This gives, from
Eq.\ (\ref{e:shuttle-dyn1}), an implicit relation between the shuttle
position and the dot occupation,
\begin{equation}
n(X)=\frac{\gamma_L e^{-4\eta X}}{\gamma_L e^{-4\eta X}+\gamma_R
e^{4\eta X}}.
\end{equation}
Near the equilibrium point, $X=0$, this implies that for
$\gamma_L=\gamma_R$, $n=0.5$. Away from the equilibrium point we
have that
\begin{equation}
n(X)=\left \{\begin{array}{ll}
0 & \ \ \  X>0 ,\\
1 & \ \ \ X<0 .
\end{array} \right .
\end{equation}
This behaviour is evident in the semiclassical dependance of $n(t)$
(thin solid line) in Fig.\ref{fig:qoshuttle}(a).

A condition for shuttling is given also by Gorelik \cite{gorelik}
by specifying the requirement for the amplitude of the shuttle
oscillation to be much bigger than the tunnelling length
$\lambda$. Donarini \cite{donarini} set the shuttling condition as
to when the mechanical relaxation rate is much smaller than the
mechanical frequency and also that the average injection and
ejection rate is approximately equal to the mechanical frequency of the
oscillator.

The quantum dynamics may be determined by solving the master
equation in the phonon number basis of the oscillator and the charge
basis for the dot.  It is necessary to truncate the phonon number
basis high enough to include the amplitude of the limit cycle.



To overcome the numerical difficulties with simulating large number
of phonon levels for the quantum case described in Sec.V later, we
choose a set of values of $\chi$ and $\eta$ which will give a rather
small limit cycle in the semiclassical approximation in
Fig.\ref{fig:qoshuttle}.  The accuracy of the semiclassical simulation
is dependent on $\lambda$ as can be seen in Sec.V by comparing
the factorized and unfactorized result from the numerical method.

We now return to consider the dependance of the total current on the
oscillator position. The total current through the device is given
by Eq.(\ref{e:averageI}). In the semiclassical approximation this is
given by
\begin{equation}
I_T(t)=\frac{\gamma_L}{2}(1-n)e^{-4\eta
X}+\frac{\gamma_R}{2}e^{4\eta X}n.
\end{equation}
At the fixed point region, this $n$ can be substituted by $n_*$
given in Eq.(\ref{e:fpn}) to give:
\begin{eqnarray}
I_T= \frac{\gamma_L\gamma_R}{\gamma_L e^{-4\eta X_*}+\gamma_R e^{4
\eta X_*}}. \label{e:Iss0}
\end{eqnarray}
When $\eta$ is small, we can simplify the current further to:
\begin{eqnarray}
I_T= \frac{\gamma_L\gamma_R}{\gamma_L +\gamma_R - 4 \eta X_*
(\gamma_L-\gamma_R)} .
\end{eqnarray}
Here we need to remember that the tunelling rates $\gamma_L$ and
$\gamma_R$ determine the steady state position of $X_*$. We can
express, from Eqs. (\ref{e:fpn}) and (\ref{e:fpx}) the tunneling
rates $\gamma_R$ as:
\begin{eqnarray}
\gamma_R = \frac{\gamma_L (B-X_*)(1-4\eta X_*)}{X_*(1+4\eta X_*)},
\end{eqnarray}
where for simplicity we have set:
\begin{eqnarray}
B = \frac{\chi v}{v^2+(\kappa/2)}.
\end{eqnarray}

We can thus rewrite the current:
\begin{eqnarray}
I_T=\frac{\gamma_L(B-X_*)}{B+4\eta X_*}. \label{e:Iss1}
\end{eqnarray}
We can see that when $\eta$ is small the current $I_T$ is linearly
dependent on the fixed point position $X_*$, with a slope of
$\frac{-\gamma_L}{B}$.

\begin{figure}[h!]
\centering
\includegraphics[width=8cm]{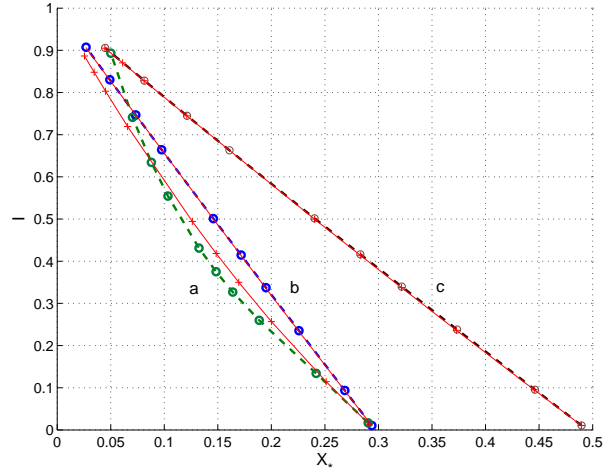}
\caption{Current versus steady state position of the oscillator for
various combination of $\eta$ and $\chi$ with varying ratio of
$\gamma_L$ and $\gamma_R$ along each curve. Here we have chosen
$\kappa=0.2$ a:$\eta=0.3,\chi=0.3$, b: $\eta=0.03,\chi=0.3$,
 c: $\eta=0.03,\chi=0.5$. Bold lines are results from the full quantum simulation
 and thin lines for the semiclassical approximations.} \label{fig:IvsX}
\end{figure}

We check this result using the full quantum simulation
(Eqs.(\ref{elect-num}), (\ref{phon-num})) and compare it with the
result of the semiclassical current Eq.(\ref{e:Iss0}).  We plot the
result for various combination of $\eta$ and $\chi$ in
Fig.\ref{fig:IvsX}. For each condition, we vary the ratio of
$\gamma_R$ to $\gamma_L$ to give the plotted curve.
As we can see from Fig.\ \ref{fig:IvsX}(b),(c), the current is indeed linearly
proportional to the steady state position of the oscillator when
$\eta$ small. In this case, the semiclassical expression of the
current given above in Eq.\ (\ref{e:Iss1}) is a very good
approximation of the actual current.


\section{A position transducer scenario}

In this section, for simplicity,  we assume that the zero
temperature limit applies for which bound state of the dot is well
below the Fermi level in the source and well above the Fermi level
in the drain. The irreversible dynamics are then conveniently
described in terms of two conditional Poisson jump processes with
rates defined in Eqs.(\ref{left_jump},\ref{right_jump}).
The jump process Eq.(\ref{left_jump}) can only occur if there are no
electrons on the dot, and the jump process Eq.({\ref{right_jump})
can only occur if there is an electron on the dot. In the case that
there is no electron on the dot, the quantum dot moves in a
quadratic potential centered on the origin. In the case that there
is an electron on the dot, the non-zero electrostatic force means
the quantum dot oscillates in a quadratic potential displaced from
the origin by $X_0=\chi/\nu$. We thus have a picture of a system
moving on one or the other potential surfaces interrupted by jumps
between them. This is schematically illustrated in
Fig.\ref{fig:jump_picture}. Due to the exponential dependance of the
jump rates on position (see Eqs.\ref{left_jump} and
\ref{right_jump}), the process $dN_L(t)$ is vastly more likely to
occur when $X<0$ and conversely, the jump process $dN_R(t)$ is much
more likely to occur when $X>0$. This means that the jump processes
are an indication of which side of $X=0$ the dot is located.

\begin{figure}[h!]
\centering
\includegraphics[width=8cm]{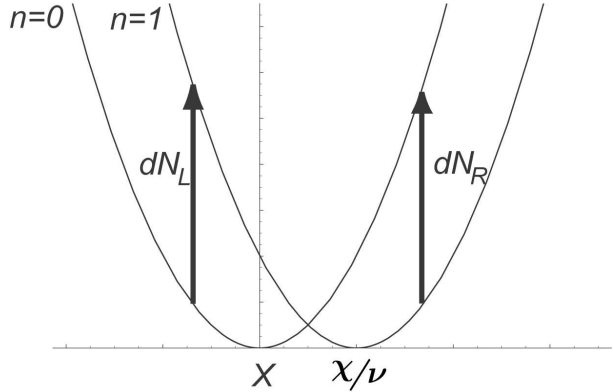}
\caption{A schematic illustration of the two potential surfaces
connected by Poisson jumps. } \label{fig:jump_picture}
\end{figure}

With this interpretation we can easily describe the conditional
dynamics of the shuttle conditioned on a history of jump processes.
In quantum optics such conditional  dynamics are called quantum
trajectories\cite{molmer,carmichael_open}.  Let us suppose that at
time $t=t_k$ , the occupation of the dot is zero and the jump
process $dN_L(t)$ occurs at $t=t_k+dt_k$. The dot then becomes
occupied while the state of the oscillator changes according to
\cite{gardiner-zoller}
\begin{equation}
|\psi(t_k)\rangle
\stackrel{dN_L}{\rightarrow}|\psi_L(t_k+dt_k)\rangle=
\frac{1}{\sqrt{p_L(\mu,t_k)}}e^{-\mu\hat{x}/2}|\psi(t_k)\rangle
\end{equation}
where $p_L(\mu,t_k)=\langle
\psi(t_k)|e^{-\mu\hat{x}}|\psi(t_k)\rangle$ and we have defined
$\mu=2/\lambda$. With these definitions we see that ${\cal
E}(dN_L(t_k))=\gamma_L p_L(\mu,t_k)dt$. We can develop some useful
insight into what this state transformation means in the case that
$|\psi(t_k)\rangle$ is a Gaussian with mean position of $\bar{x}_k$
and variance $\sigma_k$. In this Gaussian case we have
\begin{equation}
p_L(\mu,t_k)=e^{-\mu\bar{x}_k}\sum_{m=0}^\infty \frac{(\sigma_k
\mu^2)^m}{(2m)!}
\end{equation}
where $(2m)!=2.4.6....2m$. After the jump process the mean
position changes to
\begin{equation}
\langle\psi_L|\hat{x}|\psi_L\rangle=\bar{x}_k-2\sigma_k/\lambda.
\end{equation}
This equation applies equally well to jumps to the right, $dN_R$,
with a change in the sign of $\lambda$. Thus we see that if there is
jump due to $dN_L$, on average the conditional state moves to state
with a mean {\em closer} to the source, while if a jump occurs to
the right, $dN_R$, the conditional state changes to a state with a
mean position {\em closer} to the drain. This conditional behaviour
is consistent with the interpretation of the jumps as effective
measurements of the position of the quantum dot. More discussions on
the quantum trajectory picture and numerical simulations on the
conditional dynamics will be presented in the next section.

\section{Solving master equation numerically}

With the help of the Quantum Optics toolbox \cite{qotoolbox}, we
can solve the master equation directly by finding the time
evolution of the density matrix. This was done by preparing the
Liouvillian matrix in Matlab and solving the differential equation
given the initial conditions.

The expectation values for any desired quantities such as the
electron number \expect{c^\dagger c}, the phonon
number\expect{a^\dagger a}, position \expect{x} and the momentum
\expect{p} of the oscillator can be calculated by tracing the
product of this quantities with density matrix $\rho$. The result
can then be plotted against time. The same method can be applied to
calculate the steady state solution of the expectation values using
$\rho_{ss}$.

The initial state of the system has been set up to incorporate the
two electron levels, namely the occupied and empty state, combined
with an $N$ levels of phonon. The number of phonon levels included
determines the accuracy of the calculation. Of course the more
phonon levels included the more accurate the simulation will be.
However only a limited number of phonon levels can be considered.
This is due to the limited computer memory that is available and
also considering the calculation time which will be significantly
higher for larger $N$. Thus we try to find the minimum number of
phonon levels which gives convergent results. This will ensure that
the simulation still has a reasonably accurate solution. Donarini
\cite{donarini} use the Arnoldi iteration\cite{golub} to find the
stationary solution of the matrix to overcome this memory problem.
However here we have proceeded without, in the hope of looking at
not only the stationary solution but also the dynamical evolution of
the shuttle.

The behaviour of the shuttle depends strongly on the rate of
electron jump between the island and the leads. We investigate this
by looking at the variation in the electron number expectation
$\expect{c^\dagger c}$ at various rates $\gamma_L, \gamma_R$. This
is shown in Fig.\\ref{fig:3DVarGamma} in which we have set
$\gamma_L$ to be equal to $\gamma_R$. When $\gamma_L, \gamma_R$ are
small, the electron number slowly increases until it reaches the
steady state condition.
In the region where the values of $\gamma_L,\gamma_R$ is close to
the frequency of the island, oscillation starts to occur, and
depending on the damping that was set, the electron number can reach
a steady oscillation putting the system well in the oscillatory
regime. When $\gamma_L$, $\gamma_R$ are very large compared to other
frequency scales in the system, we will arrive at the strongly
damped regime of the shuttle (see Sec.\ V.B), where the jump rate of
the electron is fast enough to damp the oscillations in the electron
occupation number of the island. Since we set $\gamma_L$ to be equal to
$\gamma_R$ the steady state happens at $\expect{c^\dagger c}= 0.5$.

\begin{figure}[htbp]
\centering
\includegraphics[width=10cm]{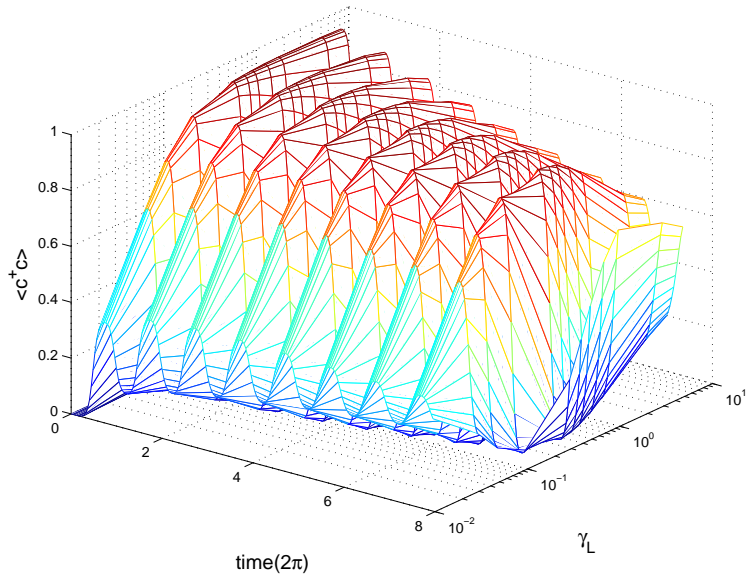}
\caption{Plot of the electron occupation number in the island for
various tunneling rate $\gamma_L$ in a logarithmic scale when
$\eta=0.3$ $\chi=0.5$, $\nu=1$ $\kappa=0.05$.}
\label{fig:3DVarGamma}
\end{figure}

\begin{figure}[htbp]
\centering
\includegraphics[width=10cm]{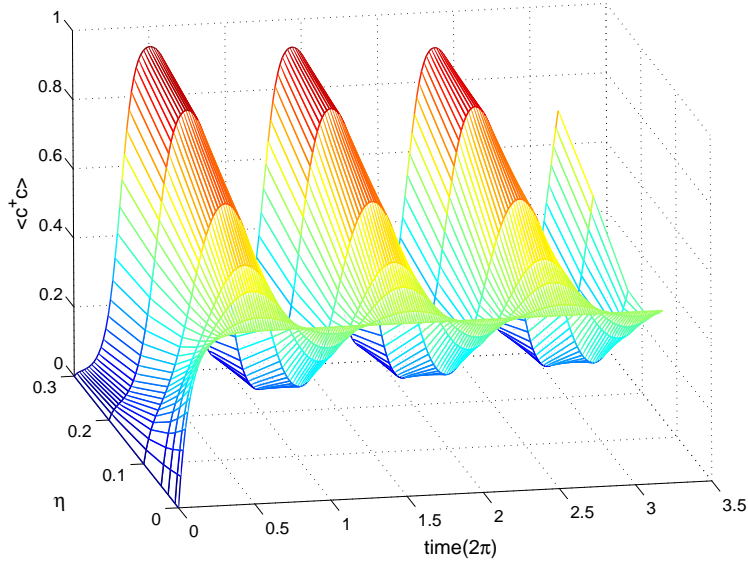}
\caption{Plot of the electron occupation number in the island for
various values of $\eta$ when $\gamma_L=\gamma_R=1$, $\chi=0.5$,
$\nu=1$ $\kappa=0.05$.} \label{fig:cdagcvareta}
\end{figure}

Similarly the behaviour of the shuttle also changes according to
$\eta$, as described in Fig.\ref{fig:cdagcvareta}. At $\eta=0$ the
electron occupation number grows to a steady state. As we increased
$\eta$ further, the oscillations start to occur with increasing
amplitude. Here we use 100 phonon levels for the numerical
calculation.

\subsection{Oscillatory Regime}

The oscillatory regime will occur when the oscillation caused by the
electron jump rate introduces continued kicks on the island. This
happens when the jump rate is close to the oscillation frequency of
the island ($\gamma_L \approx \nu$). By setting an appropriate
damping to this oscillation ($\kappa$), there exists a condition
where the island will keep oscillating between the leads. With this
setup, the system will be in the shuttle regime.

We then choose a set of parameters where the system shows the
behaviour of a shuttle, that is a continued oscillation of the
electron number along with the oscillation of the island position.
To ensure the convergence of the numerical solution, we use a
smaller value of $\eta$ that will still give a shuttling behaviour.
We choose a combination of $\eta$ and $\chi$ that will give the
smallest limit cycle to minimise the truncation error.

Within the region where the limit cycle exists, we can plot the
electron expectation number against the average position and
momentum and observe the shape of the limit cycle. We explored both
the full density matrix simulation and the semiclassical solution to
be compared.
\begin{figure*}
\subfigure[] {\includegraphics[width=9cm]{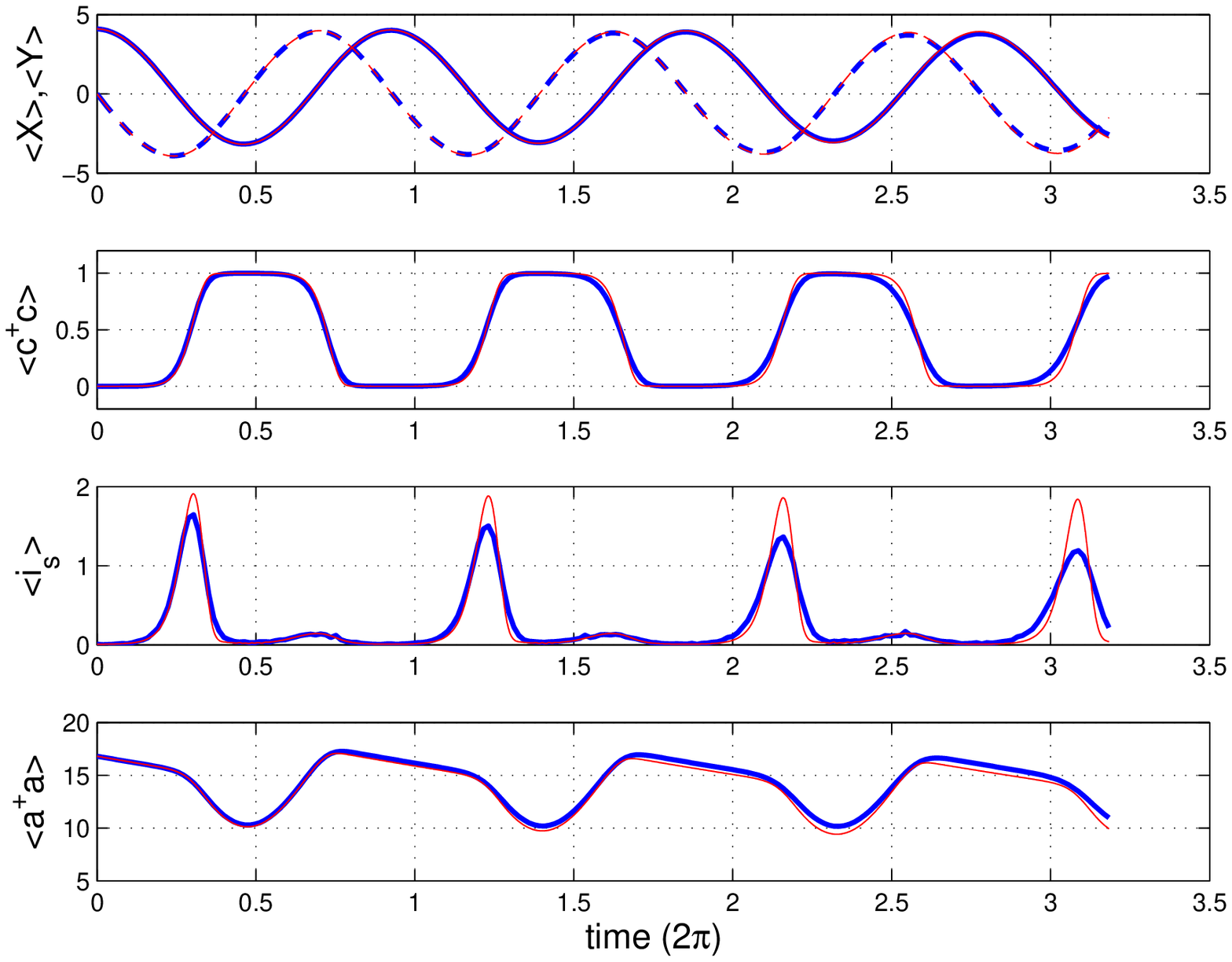} } \subfigure[]
{\includegraphics[width=5.2cm]{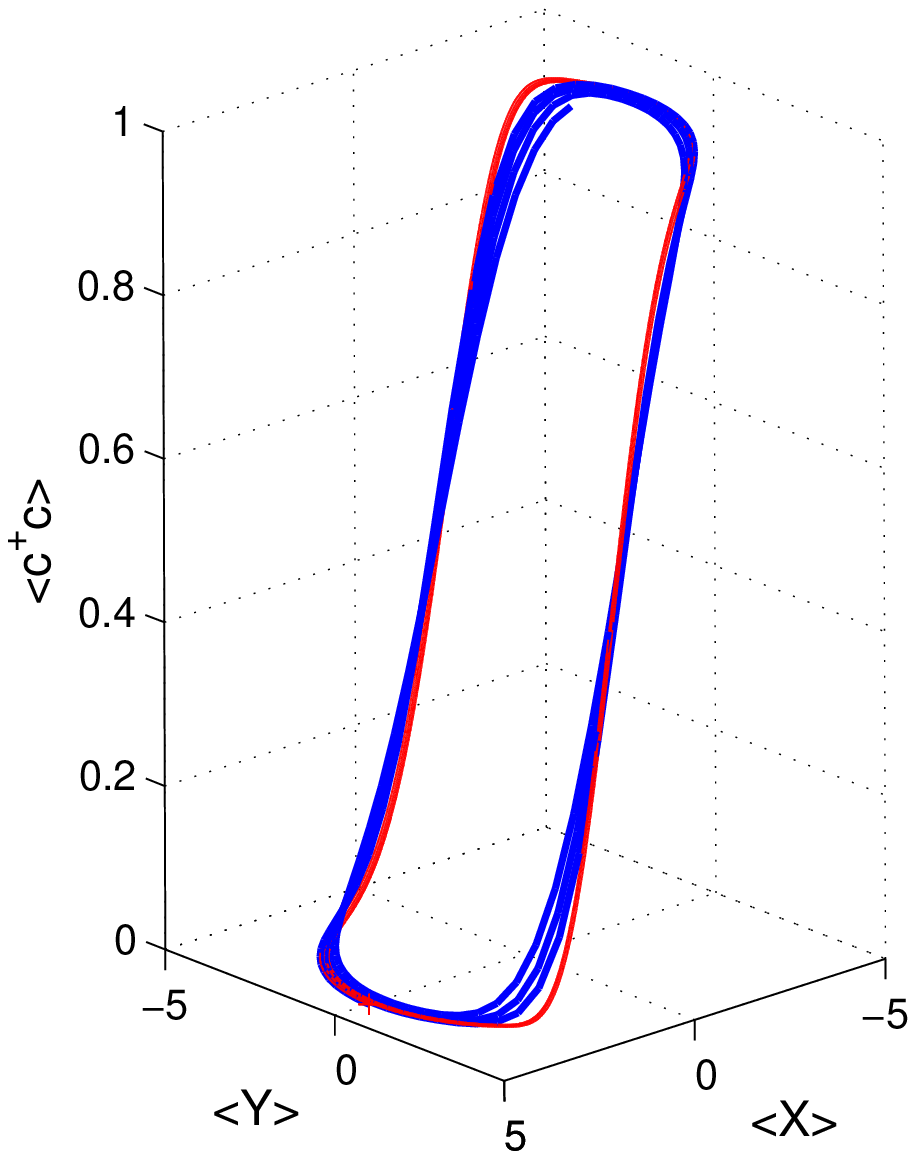} }
\caption{QO toolbox simulation for shuttling condition (bold lines),
compared with the semiclassical simulation (thin lines). Parameters
are $\gamma_L = \gamma_R = 1,\nu = 1, \eta = 0.3, \chi = 1, \kappa =
0.05.$. We set the initial condition as $X_0=4.1$ and $Y_0=0$ to
start the evolution close to the limit cycle. (a) Evolution of
average values for position (solid lines) and momentum (dashed
lines), electron number, current in the source and the energy of the
oscillator; (b) Limit cycle behaviour in 3D.} \label{fig:qoshuttle}
\end{figure*}
From the result (Fig.\ref{fig:qoshuttle}), the quantum simulation
appears to be more damped than its semiclassical counterpart.  This
is due to the effect of the noise. This slight difference can also
be caused by the dependence of the electron number on its
correlation with the position that was ignored in the semiclassical
case.

To check this we have plotted the difference between the factorized
and unfactorized moment at this particular variable combination
(Fig.\ref{fig:Factorisednun}). The time range in which the
difference in the factorized and unfactorized occurs agrees well
with the time range when the semiclassical and the quantum
simulation disagree in Fig.\\ref{fig:qoshuttle}. This disagreement
happens at the time when the shuttle is in transition between the
zero and one electron occupation number.

\begin{figure}[htbp]
\centering
\includegraphics[width=10cm]{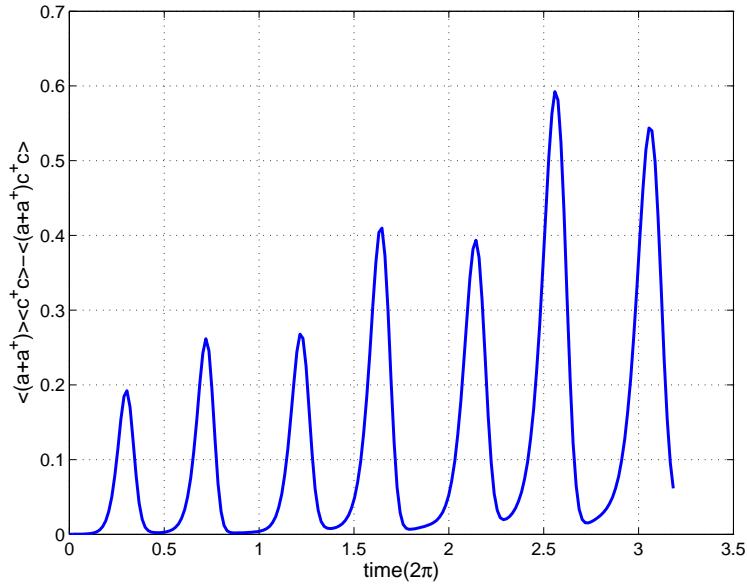}
\caption{Plot of the difference between the factorized and
unfactorized moments $\expect{(a+a^\dagger)}\expect{c^\dagger
c}-\expect{(a+a^\dagger)c^\dagger c}$ in Fig.\ref{fig:qoshuttle}.}
\label{fig:Factorisednun}
\end{figure}

Of course the truncation will pose some inaccuracy in the quantum
simulation at a longer time. However we have checked that this is
not the case at least for a short period of time by comparing it
with a simulation that includes a larger phonon number.

To investigate the effect of $\eta$ on the correlation between the
factorized and unfactorized moments, we can plot $\expect{cc^\dagger
 (a+a^\dagger)}$ and $\expect{cc^\dagger}\expect{a+a^\dagger}$.
\begin{figure}[htbp]
\centering
\includegraphics[width=10cm]{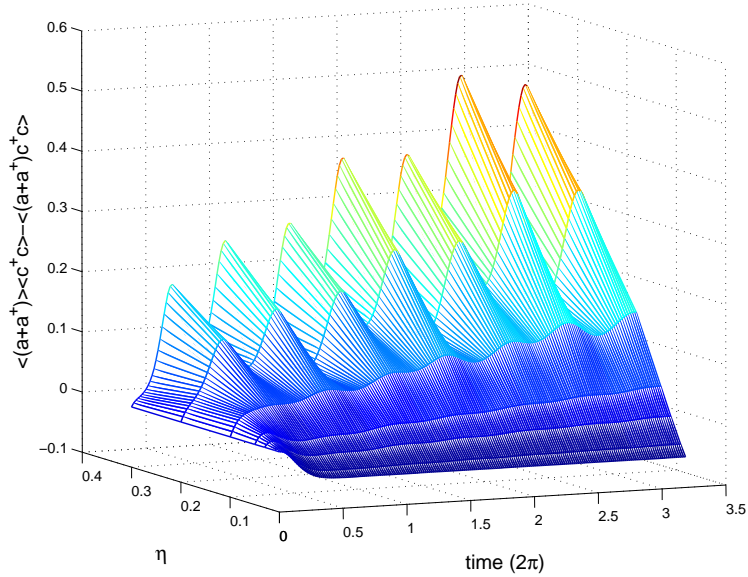}
\caption{Plot of the difference between the factorized and
unfactorized moments for various $\eta$ with $\gamma_L=\gamma_R=1,
\chi=0.5, \kappa=0.05$.} \label{fig:xcdcfactorizediffvareta}
\end{figure}
We can see from Fig.\ref{fig:xcdcfactorizediffvareta}, that the
semiclassical approximation agrees with the quantum simulation under
the condition that $\eta$ is small enough. As $\eta$ increases, the
evidence of this difference becomes noticeable. This difference
oscillates with peaks located at times when electron jumps happen,
that is when the oscillator is near the equilibrium position.


As opposed to what the name "electron shuttle" suggests,  the
dynamics in the shuttle regime, for the parameters specified in
Fig.\ref{fig:qoshuttle}, is not like a conventional shuttle which
picks up an electron when it is  closest to the source and
drops the electron when it is closest to the drain, as also
suggested by Nord \ea \cite{nord}.
Looking at the rate of the average electron number and the average
current in the source (Fig.\ref{fig:qoshuttle}(a)), this is
certainly not the case. The shuttle picks up an electron near an
average displacement of zero, slightly towards the source, and
continues to travel closer to the source electrodes. It then
oscillates back and drops the electron at a slightly displaced
average position from equilibrium towards the drain electrode. The
shuttle then continues to get closer to the drain before oscillating
back to repeat the cycle.

An important distinction must be made between the dynamics of the
averages derived from solving the master equation and a dynamics
conditioned on a particular history of tunneling events. This
distinction is already suggested by inspecting the average electron
number as a function of time. In any actual realisation of  the
stochastic process, the number of electrons on the dot is either
zero or one, yet the ensemble average occupation number varies
smoothly between zero and one.  The reason for this is that the
actual times at which transitions between the two states takes place
fluctuates.

We can more easily appreciate this distinction using an alternative
approach to understanding the dynamics based on `quantum
trajectories'.  The quantum trajectory method (sometimes called the
Monte Carlo method) first introduced in quantum optics, is a method
of looking at the evolution of a system conditioned on the results
of measurements made on that system. \cite{dum, carmichael_open,
goan_continuous, goan_dynamics, goan_montecarlo}. This method will
allow one to monitor 'events' such as the jump of an electron to the
island which causes the displacement kick on an oscillator.

The Quantum Optics Toolbox enables a direct computation of the
conditional dynamics of the operator moments by implementing a so
called 'jump unravelling' of the master equation. First we plot a
sample trajectory for a slow electron jump rate $\gamma_L=0.1$ to
see the effect of electron jump on the evolution picture of the
system. A random jump of electrons from the source to the island
(Fig.\ref{fig:SingleTrajlambdap3}) according to rate $\gamma_L,
\gamma_R$ was introduced. The dynamics of the shuttle as a position
transducer, as predicted in section IV, can be seen in the
conditional averages of the displacement. The variable $\eta$
controls the amount of displacement of the island when an electron
jumps on and off onto the island. Larger value of $\eta$ caused a
larger displacement kick when a jump occurs. During the time when
the electron is on the island, the phonon number of the oscillator
oscillates with a similar behaviour to the oscillation of the
position.

\begin{figure}[htbp]
\centering
\includegraphics[width=12cm,angle=0]{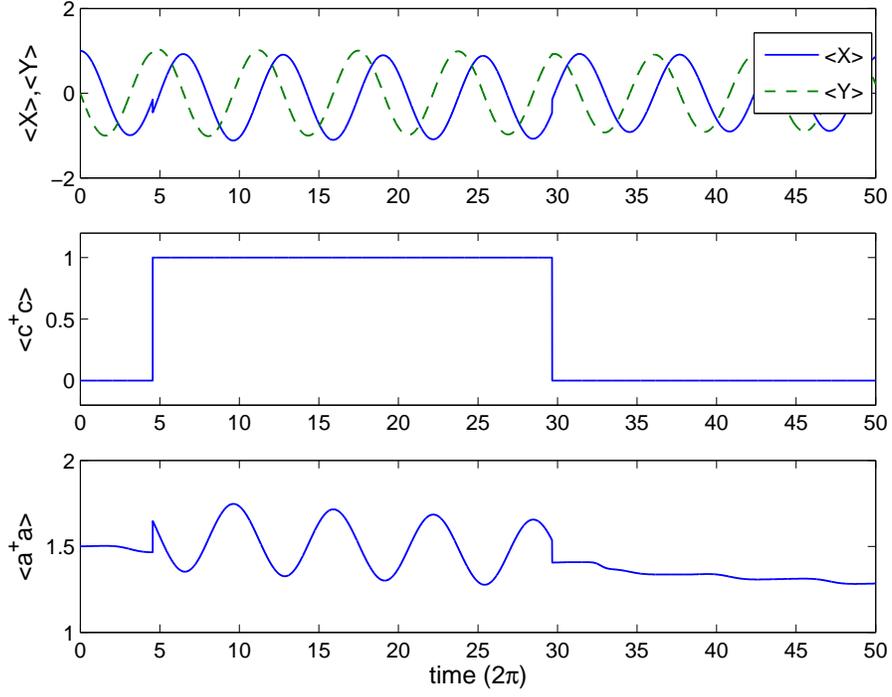}
\caption{Plot of a single trajectory showing the dynamics of the
jump and the result in the phonon number. Parameters are $\gamma_L =
\gamma_R = 0.1, \nu = 1, \eta = 0.3, \chi = 1, \kappa = 0.05$.}
\label{fig:SingleTrajlambdap3}
\end{figure}

\begin{figure}[htbp]
\centering
\includegraphics[width=12cm]{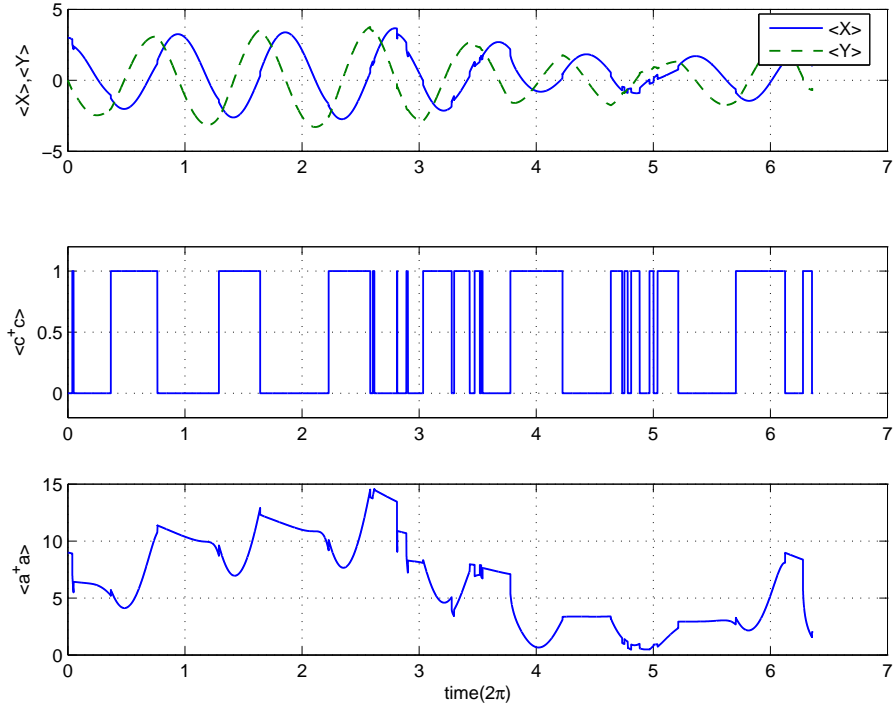}
\caption{Plot of a single trajectory showing the dynamics of the
jump when $\gamma_L = \gamma_R = 1, \nu = 1, \eta = 0.3, \chi = 1,
\kappa = 0.05.$} \label{fig:singletraj_gammaL1_lambdap3_Ec1permak}
\end{figure}
The single trajectory for the shuttle case with the same parameter
in Fig.\ref{fig:qoshuttle} can be seen in
Fig.\ref{fig:singletraj_gammaL1_lambdap3_Ec1permak}. The electrons
mostly jump onto the island from the source when it is closer to the
source and jump off when closer to the drain. At the jump, the
island gets a slight displacement kick towards the source when
jumping on and towards the drain when jumping off. However this does
not stop the shuttling motion of the island and does not repel it to
the opposite direction as suggested earlier by Nord \ea \cite{nord}.
It can also be seen that when the electron manages to jump onto the
island when island is still close to the drain, it is more probable
for the electron to jump off straight away.

The conditional dynamics of the system just described corresponds to
an experiment in which number of electrons on island is monitored
continuously in time. As we can see, the behaviour of the
conditional dynamics differs from  the behaviour of the ensemble
average. However, averaging over many different realization of the
trajectories as shown in
Fig.\ref{fig:singletraj_gammaL1_lambdap3_Ec1permak} would lead to a
closer and closer approximation of the ensemble average behaviour in
Fig.\ref{fig:qoshuttle}.

\subsection{Strongly Damped Regime}

There are two ways of damping the shuttle into the fixed point
regime. One is to damp the motion of the shuttle itself by
introducing a large mechanical damping $\kappa$. Alternatively we
can damp the oscillation of the electron occupation number in the
island. This happens when the rates of the electron jump
$\gamma_L,\gamma_R$ are large compared to the natural frequency of
the island vibration. The fast electron jumps act as an internal
damping to the shuttle. Within this regime the electron number
expectation \expect{c^\dagger c} monotonically approaches 0.5 when
$\gamma_L = \gamma_R$.

When the bare electron tunnelling rates are very large compared to
other frequency scales in the problem, we may assume the dot
approaches its steady state for bare tunnelling quickly as
compared to the typical time scale of the oscillator. In this case,
$\rho(t)=\rho_d^{\rm st} \times \rho_o(t)$. The bare $\rho_d^{\rm
st}$ can be substituted into the density matrix of the total master
equation and then be traced out with respect to the dot degrees of
freedom to get an effective master equation which involves only the
reduced density matrix of the oscillator. This effective master
equation can be calculated from the reduced density matrix, from Eq.
(\ref{e:Imastereqn}).

Since $\eta$ is assumed to be small, we can expand the expression to
second order in $\eta$: $e^{\eta \hat{X}} = 1 + \eta \hat{X} + (\eta
\hat{X})^2/(2!)+\cdots$. We can then re-write the zero temperature
full master equation as:
\begin{eqnarray}
\dot{\rho} &=& -i \nu [a^\dagger a, \rho] \nonumber \\
&+& 2i \chi [\hat{X} \bar{n}, \rho] \nonumber \\
&+& 2 \gamma _{L} (1-\bar{n}) \eta^2 [\hat{X},[\hat{X},\rho]] \nonumber\\
&+& 2 \gamma _{R} \bar{n} \eta^2 [\hat{X},[\hat{X},\rho]] \nonumber\\
&+& \kappa(\bar{n}_p+1) \mathcal{D} [a] \rho + \kappa \bar{n}_p
\mathcal{D} [a^\dagger] \rho, \label{e:mastereqn_dottrace}
\end{eqnarray}
with $\bar{n}=\gamma_L/\gamma_L+\gamma_R$. The following moments can
thus be derived from Eq.\ (\ref{e:mastereqn_dottrace}):
\begin{eqnarray}
\frac{d\expect{a^\dagger a}}{dt} &=& 2 \chi \bar{n} \expect{\hat{Y}}
- 4 \gamma_L(1-\bar{n})\eta^2
-4 \gamma_R \bar{n}\eta^2  + \kappa \bar{n} - \kappa \expect{a^\dagger a}\\
\frac{d\expect{\hat{X}}}{dt} &=& \nu \expect{\hat{Y}} - \frac{\kappa}{2} \expect{\hat{X}} \\
\frac{d\expect{\hat{Y}}}{dt} &=& -\nu \expect{\hat{X}} + \chi
\bar{n} - \frac{\kappa}{2} \expect{\hat{Y}}.
\end{eqnarray}

The moments $\expect{\hat{X}}$ and $\expect{\hat{Y}}$ form a closed
system of differential equation which can readily be solved.
\begin{eqnarray}
\expect{\hat{X}} &=& e^{-(\kappa /2)t} \biggl( (X_0 - X_*)\cos(\nu
t)+ (Y_0 - Y_*) \sin(\nu t) \biggr)+X_*, \label{e:strongX}\\
\expect{\hat{Y}} &=& e^{-(\kappa /2)t} \biggl( - (X_0 - X_*)\sin(\nu
t)+ (Y_0 - Y_*) \cos(\nu t)\biggr)+Y_*.\label{e:strongY}
\end{eqnarray}
where again $X_*$ and $Y_*$ is simply the displacement in the
equilibrium such as given in Eqs.(\ref{e:fpx}) and (\ref{e:fpy})
with $n_*=\bar{n}$. $X_0$ and $Y_0$ is the initial condition of $X$
and $Y$ respectively.  The analytic expressions of
Eqs.(\ref{e:strongX}) and (\ref{e:strongY}) are useful for checking
the solution of the master equation given by Matlab, to ensure that
the truncation in the phonon number is adequate.

\begin{figure}[h!]
\centering
\includegraphics[width=8cm]{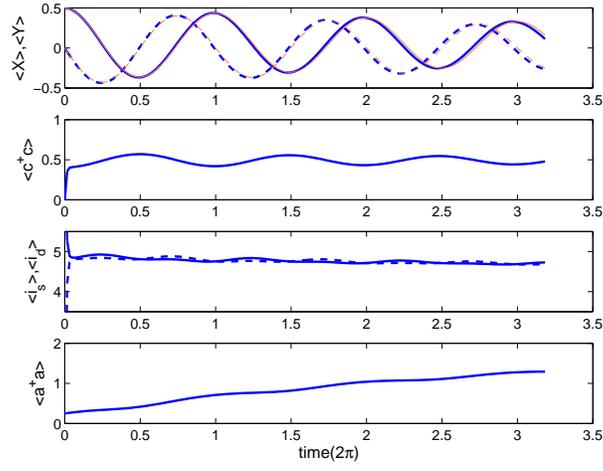}
\caption{Average evolution of the shuttle with a large $\gamma_L$
and $\gamma_R$. Here we have chosen $\gamma_L = \gamma_R = 10,
\eta=0.1, \chi=0.1, \kappa=0.05, X_0=0.5, Y_0=0$ for both quantum
simulation with $N=25$ (thick lines) and semiclassical solution
(thin lines).} \label{fig:SingleTrajStrongDamp}
\end{figure}

The shuttle oscillation is damped to the new displaced position of
$X_*$ which agrees to the obtained result previously. When
$\gamma_L=\gamma_R=gamma$, we have $n_* = \bar{n} = \half$ in the
regime when the tunneling rates are very large compared to other
frequency scales (especially when $\eta$ is relatively small). In
this case, the oscillatory behaviour of Eqs.\ (\ref{e:strongX}) and
(\ref{e:strongY}) do not depend on the actual values of the
tunneling rates $gamma$. It can also be deduced that the decay rate of the
oscillation envelope is $e^{-\kappa t/2}$. In this regime, the
result of the analytical expressions matches the quantum simulation
quite well.

\subsection{Co-existence regime}

As discussed in Sec. III, we can also have a regime in which the
behaviour of the shuttle depends on its initial condition. The
system will either be attracted to the limit cycle and thus be in
the shuttle regime or be attracted to the fixed point and be in the
tunneling regime depending on its initial condition within the
correct parameters where the subcritical bifurcation occurs.
Following previous authors\cite{donarini}, we call this the
'co-existence regime'.

Semiclassically this can be seen when we plot the average evolution
of the shuttle. Depending on the initial conditions, the shuttle
will either be attracted to the fixed point position or undergoes
the stable limit cycle oscillation (Fig.\ \ref{fig:ssc_limit}).

\begin{figure}[h!]
\centering
\includegraphics[width=8cm]{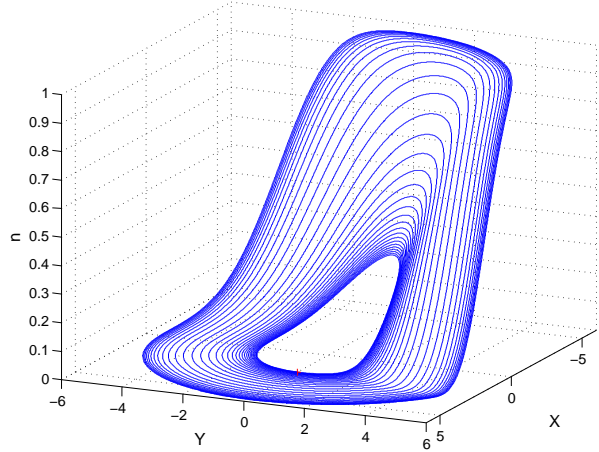}
\caption{Semiclassical limit cycle of the shuttle in the
co-existence regime. Here we have chosen $\gamma_L = 0.03, \gamma_R
= 1, \eta=0.3, \chi=1, \kappa=0.05, X_0=2.2$ and $Y_0=0$. The
evolution starts close to the unstable limit cycle and then moves
toward the stable limit cycle} \label{fig:ssc_limit}
\end{figure}

The quantum average calculation in this regime however does not show
the subcritical bifurcation since averaging over the noise in the
system dampens this effect. This can be seen in the evolution of the
single trajectory which is captured to the fixed point position at
random times. A sample of trajectories each from different initial
conditions were plotted in figure \ref{fig:SingleTraj_fixed} and
\ref{fig:SingleTraj_jump}.

\begin{figure}[h!]
\centering
\includegraphics[width=10cm]{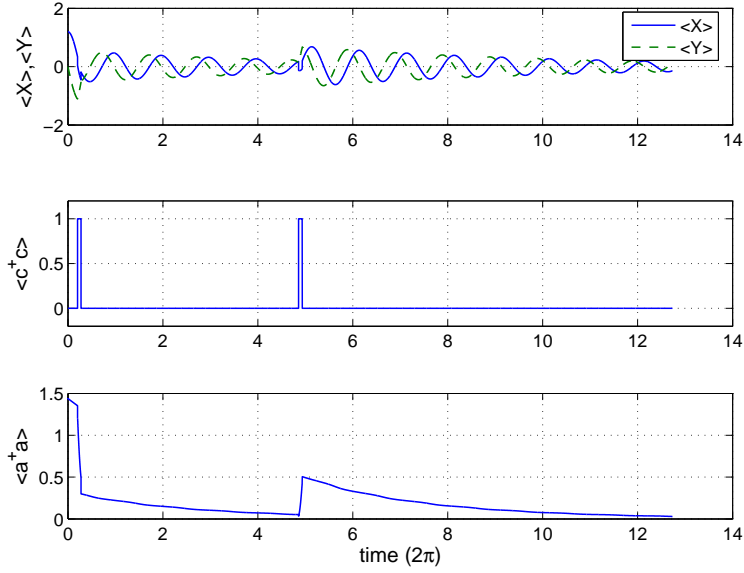}
\caption{Average evolution of the shuttle.
Here we have chosen $\gamma_L = 0.01, \gamma_R = 1,
\eta=0.3, \chi=1, \kappa=0.05, X_0=1.2, Y_0=0$.}
\label{fig:SingleTraj_fixed}
\end{figure}

\begin{figure}[h!]
\centering
\includegraphics[width=10cm]{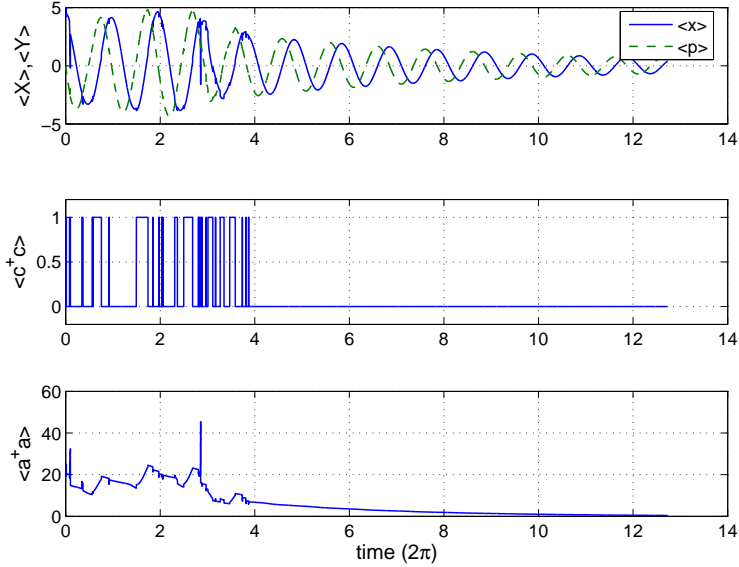}
\caption{Average evolution of the shuttle.
Here we have chosen $\gamma_L = 0.01, \gamma_R = 1,
\eta=0.3, \chi=1, \kappa=0.05, X_0=5, Y_0=0$.}
\label{fig:SingleTraj_jump}
\end{figure}

\subsection{Finite Temperature}

We can easily extend these calculations to the finite temperature
case by including the fermi factor $f_L$ and $f_R$, which was
previously set to 1 and 0 for the zero temperature case, in the
calculation for both the full quantum simulation and the
semiclassical approximation.

\begin{figure}[h!]
\centering
\includegraphics[width=10cm]{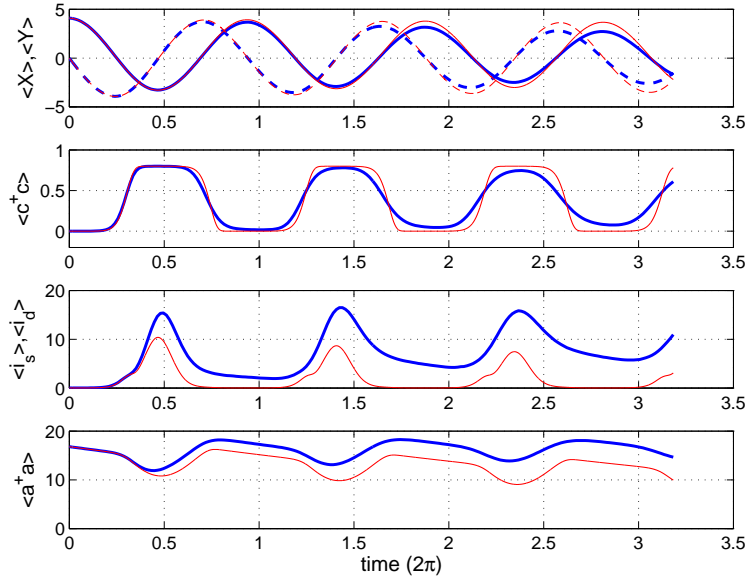}
\caption{Average evolution of the shuttle with finite Temperature
$\gamma_L$ and $\gamma_R$. Here we have chosen $\gamma_L = \gamma_R
= 1, \eta=0.3, \chi=1, \kappa=0.05, X_0=4.1, Y_0=0, f_L=0.8$ and
$f_R=0$ at $\omega_I = 3\nu$ for both quantum simulation with $N=47$
(thick lines) and semiclassical approximation (thin lines).}
\label{fig:SingleTrajStrongDamp}
\end{figure}

The effect on temperature on the system is shown in
Fig.\ \ref{fig:SingleTrajStrongDamp}.  Comparing this with previous
result for the zero temperature (Fig.\ \ref{fig:qoshuttle}) there is a
suppression of the electron number oscillation. There is also a
significant difference between the quantum and the semiclassical
simulation for the electron number occupation which resulted in a
difference in the current in each of the leads.

\section{Noise calculation}
In surface gated 2DEG structures  some recent experiments monitor
the charging state of the dot via conductance in a quantum point
contact\cite{kouwenhoven}. However such techniques cannot easily be
adapted for a nanoelectromechanical system. In experiments involving
tunneling through a double barrier quantum dot structure the
simplest thing to measure is the source-drain current. In the QEMS
experiments of Park et al\cite{park} and also of Erbe et al, the
source drain current carried signatures of the vibration of the
nanoelectromechanical component.  In this section we calculate,
using the Quantum Optics Toolbox,  the current noise spectrum and
show that it indicates the transition between the fixed point and
the shuttling regime.

The current seen in the external circuit, when electrons tunnel on
and off the dot, only indirectly reflects the quantum nature of the
tunneling process.  Tunneling causes a local departure from
equilibrium in the source and drain reservoirs that is restored
through a fast irreversible process in which small increments of
charge are exchanged with the external circuit. While tunneling
obviously involves a change of charge in units of $\pm$e, the
increments of charge drawn by the external circuit are continuous
quantities determined by the overall capacitance and resistance of
the circuit. The current responds as a classical stochastic process
conditioned on the quantum stochastic processes involved in the
tunneling.  In many ways this is analogous to the response of a
photo electron detector to photons.

The connection between the quantum stochastic process of tunneling
and the current observed in the external circuit is given by the
Ramo-Shockley theorem and is a linear combination of the two Poisson
processes defined in Eqs.\ (\ref{left_jump},\ref{right_jump}).  The
noise spectrum of such a current involves moments of both the
tunneling processes, and correlations between them. In Sun et
al.\cite{sun}, one can find a detailed example of how such
correlations are determined by the corresponding master equation for
the quantum dot system.

Recently Flindt et al.\cite{flindt}, have calculated a noise
spectrum for the shuttle system defined in terms of the fluctuating
electron number accumulating in the drain reservoir. Here we adopt a
different (but equivalent) approach based on the framework of
quantum trajectories.   In  this section we calculate, using quantum
trajectory methods, the stationary current noise spectrum in the
{\em source}  current alone as this suffices to illustrate how the
current noise spectrum reflects the transition from fixed point to
shuttling.  The total current shows the same features but has a
different noise background.

The two time correlation function quantifies the fluctuations in the
observed current and is defined by:
$$
G(\tau) = \frac{e}{2} i_\infty \delta(\tau)+ E (I(t) I(t+\tau))
^{\tau \neq 0}_{t \rightarrow \infty},
$$
The first term is responsible for shot noise in the current, while
the second term quantifies noise correlations. We now show how the
second term can be defined in terms of the stationary state of the
quantum dot itself.

Let $\rho(t)$ be the density operator representing the dot at time
$t$. What is the conditional  probability that,  given an electron
tunnels onto the dot from the drain between $t$ and $t+dt$, another
similar tunneling event takes place a time $\tau$ later (with no
regard for what tunneling events have occurred in the mean time)? If
an electron tunnels onto the dot from the drain at time $t$, the
conditional state of the dot (unnormalised), conditioned on this
event is given by
\begin{equation}
\tilde{\rho}^{(1)}(t)=\gamma_L e^{-\hat{x}/\lambda}c^\dagger
\rho(t)c e^{-\hat{x}/\lambda}.
\end{equation}
Given this state, the probability that another tunneling event takes
place a time $\tau$ later is
\begin{eqnarray*}
G(t,\tau) & = & \gamma_L{\rm tr}\left (e^{-2\hat{x}/\lambda} cc^\dagger  e^{{\cal L}\tau}[\tilde{\rho}^{(1)}(t)]\right )\\
  & = &  \gamma^2_L{\rm tr}\left (e^{-2\hat{x}/\lambda} cc^\dagger e^{{\cal L}\tau}[e^{-\hat{x}/\lambda}c^\dagger \rho(t)ce^{-\hat{x}/\lambda}]\right ).
\end{eqnarray*}
where formally we have represented the irreversible dynamics from
time $t$ to $t+\tau$ as the propagator $e^{{\cal L}\tau}$. Let us
now assume that the first conditioning event takes place at a time
$t$ long after any information about the initial state of the
quantum dot has decayed away. That is to say the first conditioning
event occurs when the dot has settled into the stationary state,
$\rho_\infty=\lim_{t \rightarrow \infty}\rho(t)$. The stationary
two-time correlation function for the source current is then defined
by
\begin{equation}
G(\tau)=\gamma^2_L{\rm tr}\left (e^{-2\hat{x}/\lambda}  cc^\dagger e^{{\cal L}\tau}[e^{-\hat{x}/\lambda}c^\dagger \rho_\infty c e^{-\hat{x}/\lambda}]\right )
\end{equation}
In terms of the dimensionless position operator, $X$, the noise in
the  two time correlation functions becomes
$$
G(\tau)=E(I_L(t)I_L(t+\tau))^{\tau>0}_{t\to\infty}= \gamma_L^2 {\rm
Tr}[e^{-4 \eta \hat{X}}cc^\dagger   e^{{\cal L}\tau}(e^{-2 \eta
\hat{X}}c^\dagger \rho_\infty c e^{-2 \eta \hat{X}})]
$$
where $e^{{\cal L}\tau}$ is the master equation evolution.

The noise power spectrum of the current is given by:
\begin{eqnarray}
S(\omega) = 2 \int_0^\infty
G(\tau)(e^{i\omega\tau}+e^{-i\omega\tau})d\tau
\end{eqnarray}
This noise spectrum can be directly calculated using the Quantum
Optics Toolbox by first calculating the steady state solution
$\rho_{\infty}$ and setting $e^{-2 \eta \hat{X}}c^\dagger
\rho_\infty c e^{-2 \eta \hat{X}}$ as an initial condition for the
master equation evolution. Then we can calculate the expectation
value of the operator $cc^\dagger e^{-4 \eta \hat{X}}$ in the state
evolved, according to the master equation, from this initial
condition.  It is important to note that the master equation does
indeed have a steady state even in that parameter regime in which
the semiclassical dynamics would imply a limit cycle. This is
because quantum fluctuations cause a kind of phase diffusion around
the limit cycle. These quantum fluctuations are precisely the random
switchings observed in the single quantum trajectory shown in figure
\ref{fig:singletraj_gammaL1_lambdap3_Ec1permak}. In fact as shown in
\cite{novotny} the Wigner function of the steady state has support
on the entire limit cycle.

The example of the noise spectra for various $\eta$ is shown in
Fig.\ref{fig:corrspectrum}.
\begin{figure}[htbp]
\centering
\includegraphics[width=9cm, angle=-90]{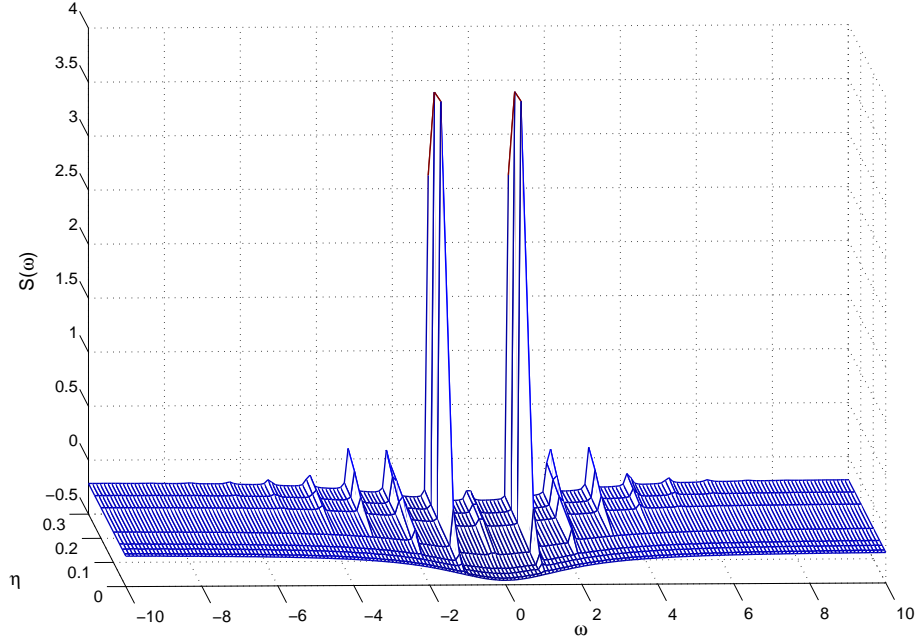}
\caption{Plot of spectrum in the left junction with various $\eta$,
setting $\chi=0.5$.} \label{fig:corrspectrum}
\end{figure}
The transition between the tunnelling regime and the shuttling
regime is clearly evident in the noise power spectrum, figure
\ref{fig:corrspectrum} (We have subtracted off the shot noise
background).

Setting $\eta=0, \chi=0$ we arrived back at the known spectrum for
the source current in a double barrier device\cite{sun} with a
single dip at zero frequency. As the values of $\eta$ and $\chi$ (or
$\gamma_L$) are increased, the frequency spectra develop sidebands
which correspond to the frequency of the oscillator
(Fig.\\ref{fig:corrspectrum},\ref{fig:corrspectrumdiffgamma}). As
the system approaches the shuttling regime, the frequency spectra
pick up noise peak at zero frequency and additional peaks at higher
frequencies close to a multiple of the oscillator frequency. This is
a signature of the limit cycle formation. On the limit cycle, the
frequency is shifted from the base oscillation frequency $\omega =
1.055 \nu $. This is also given by the imaginary part of the
eigenvalues of the linearised matrix expressed
$\mu=\sqrt{A_*\kappa+1+(\kappa/2)^2}$.  This observation agrees with
the predicted slight re-normalization of the frequency by Flindt \ea
\cite{flindt}.

\begin{figure}[htbp]
\centering
\includegraphics[width=10cm]{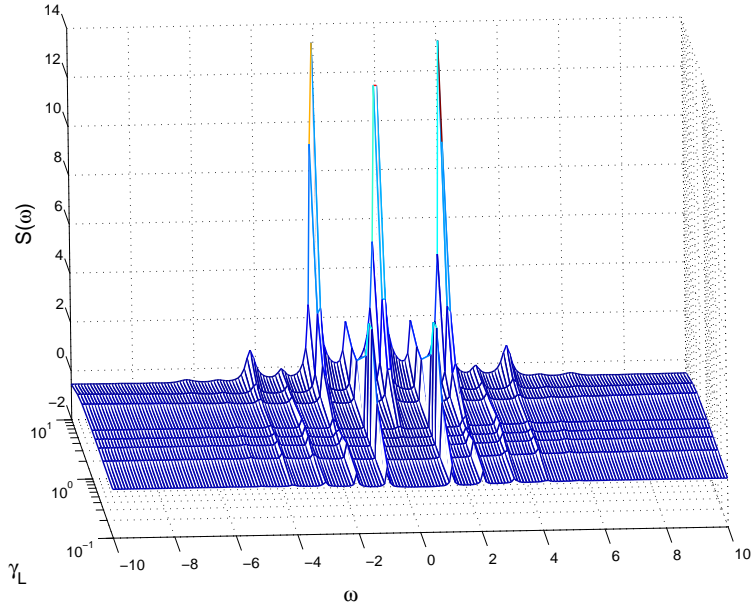}
\caption{Plot of spectrum in the left junction with various
$\gamma_L$, setting $\eta=0.3,\chi=0.5, \kappa=0.05.$ }
\label{fig:corrspectrumdiffgamma}
\end{figure}

A similar feature of the noise is found by Armour
\cite{armour_current} in a system consisting of a SET that is
coupled to a nanomechanical resonator. Although this is a
different system from the shuttle system, the classical spectrum
noise in this system also shows the dependency of the current on
the position of the nanomechanical resonator.

\section{Conclusion}

The dynamics of the shuttle system has been investigated via both
the semi-classical and the full quantum master equation treatment.
The latter reveals subtle properties of the dynamics which was not
found using the semiclassical treatment.  The master equation is
solved numerically using the Quantum Optics Toolbox enabling a
detailed comparison of the semiclassical dynamics with the quantum
ensemble averages. For the first time in the study of the quantum
shuttle  we compute the moments for the quantum state conditioned on
a particular history of tunnelling events. This is called a quantum
trajectory and it reflects what can be observed experimentally by
monitoring the electron on the island.

The conditional dynamics differs from the behaviour of the ensemble
average, and gives new insight into the shuttling dynamics. In the
shuttling regime, the ensemble average dynamics of the electron
occupation number is a smoothed square wave that slowly decays to a
steady state value of one half.  Given that the occupation number of
the dot is either zero or unity this ensemble averaged behaviour may
seem unexpected. However looking at the occupation number in  a
single conditional state (see Fig.\ \ref{fig:SingleTrajlambdap3})
indicates what is going on. A single quantum trajectory shows that
the average occupation number is indeed either zero or unity and in
the shuttling regime behaves like a square wave for short times but,
at random times, suffers a phase jump. The ensemble average of many
such trajectories with phase jumps at random times leads to the
observed ensemble average dynamics as computed from the master
equation. These random phase jumps ultimately lead to a steady state
density operator for the system that, in the Wigner representation,
is diffused around the limit cycle, as noted by Novotny et
al.\cite{novotny}.

The shuttle dynamics was investigated in two regimes: the fixed
point and the shuttle regime. In the fixed point regime, the shuttle
is damped to a new displaced position. We have shown that there is a
strong relation between the current and the fixed point of the
position. This relationship is linear when the tunnel length is
large ($\eta$ small). Thus it is possible to use the shuttle in a
position transducer scenario. In this regime, the semiclassical
treatment is shown to be accurately sufficient to describe the
dynamics.

We provide the condition in which the shuttle regime will appear
from the system by identifying the appearance of limit cycle in the
phase space of the shuttle. A careful analysis of the nonlinear
dynamics using centre manifold method indicates that when
$\gamma_L=\gamma_R$, the limit cycle forms through a supercritical
pitchfork bifurcation. However when $\gamma_L\neq\gamma_R$ there is
a region of parameter space in which the bifurcation can be
subcritical, and for which hysteresis is possible.    Adjusting the
damping $\kappa$ with respect to these parameters will cause the
shuttle to be sufficiently damped and thus allow the shuttling to
take place. The shuttle regime also appears when the rate of the
electron tunnelling is close to the oscillator frequency. The
shuttle regime corresponds to
 the continuous oscillation of the electron number and
results in additional peaks at multiples of the limit cycle
frequency in the noise spectra. This is destroyed
when $\kappa$ is too large or when a large electron jump $\gamma_L,
\gamma_R$ are introduced to the system. Both of these conditions
will damp the shuttle into the displaced equilibrium position.  The
quantum shuttle thus provides a fascinating example of a quantum
stochastic system in which electron transport is coupled to
mechanical motion. In future studies we will investigate how such a
system can be configured for sensitive force detection.

\section*{Acknowledgment}
\label{sec:Acknow}
HSG is grateful to the Centre for Quantum Computer Technology at the
University of Queensland for their hospitality during his visit. HSG
would also like to acknowledge the support from the National Science
Council, Taiwan under Contract No. NSC 94-2112-M-002-028, and
support from the focus group program of the National Center for
Theoretical Sciences, Taiwan under Contract No. NSC
94-2119-M-002-001. GJM acknowledges the support of the Australian
Research Council through the Federation Fellowship Program.

\bibliographystyle{prsty}
\bibliography{phd}

\begin{thebibliography}{10}

\bibitem{knobel-cleland}
R.~G. Knobel and A.~N. Cleland, Nature {\bf 424},  291  (2003).

\bibitem{lahaye}
M.~D. LaHaye, O. Buu, B. Camarota, and K. Schwab, Science {\bf 304},  74
  (2004).

\bibitem{ekinci}
K.~L. Ekinci, X.~M.~H. Huang, and M.~L. Roukes, Appl. Phys. Lett. {\bf 84},
  4469  (2004).

\bibitem{gorelik}
L. Gorelik {\it et~al.}, Phys. Rev. Lett. {\bf 80},  4526  (1998).

\bibitem{qotoolbox}
S. Tan, Quantum Optics Toolbox,
  http://www.phy.auckland.ac.nz/Staff/smt/qotoolbox/download.html.

\bibitem{molmer}
K. Molmer, Y. Castin, and J. Dalibard, Journal of the Optical Society of
  America B (Optical Physics) {\bf 10},  524  (1993).

\bibitem{shekhter}
R.~I. Shekhter {\it et~al.}, J. Phys.: Condens. Matter {\bf 15},  R441  (2003).

\bibitem{park}
H. Park {\it et~al.}, Nature {\bf 407},  57  (2000).

\bibitem{zhitenev}
N. Zhitenev, H. Meng, and Z. Bao, Phys. Rev. Lett. {\bf 88},  226801  (2002).

\bibitem{erbe}
A. Erbe, C. Weiss, W. Zwerger, and R. Blick, Phys. Rev. Lett. {\bf 87},  096106
   (2001).

\bibitem{huang}
X.~M.~H. Huang, C.~A. Zorman, M. Mehregany, and M.~L. Roukes, Nature {\bf 421},
   496  (2003).

\bibitem{isacsson_noise}
A. Isacsson and T. Nord, Europhys. Lett. {\bf 66},  708  (2004).

\bibitem{novotny_shot}
T. Novotny, A. Donarini, C. Flindt, and A.-P. Jauho, Phys. Rev. Lett {\bf 92},
  248302  (2004).

\bibitem{flindt}
C. Flindt, T. Novotny, and A.-P. Jauho, Physica E {\bf 28},  in press  (2005).

\bibitem{isacsson}
A. Isacsson, Phys. Rev. B {\bf 64},  035326  (2001).

\bibitem{donarini}
A. Donarini, cond-mat/0501242 v1, 2005.

\bibitem{aji}
V. Aji, J.~E. Moore, and C.~M. Varma, APS Meeting Abstracts  17004  (2003).

\bibitem{mccarthy}
K.~D. McCarthy, N. Prokof'ev, and M.~T. Tuominen, Phys. Rev. B {\bf 67},
  245415  (2003).

\bibitem{fedorets_shuttle}
D. Fedorets, L. Gorelik, R. Shekhter, and M. Jonson, Phys. Rev. Lett {\bf 92},
  166801  (2004).

\bibitem{novotny}
T. Novotny, A. Donarini, and A.-P. Jauho, Phys. Rev. Lett. {\bf 90},  256801
  (2003).

\bibitem{jauho}
A.-P. Jauho, T. Novotny, A. Donarini, and C. Flindt, cond-mat/0411107 v1  .

\bibitem{armourandmackinon}
A. Armour and A. MacKinnon, Phys. Rev. B {\bf 66},  035333  (2002).

\bibitem{gardiner-zoller}
C.~W. Gardiner and P. Zoller, {\em Quantum Noise}, 2nd ed. (Springer-Verlag,
  Berlin, 2000).

\bibitem{wahyu}
D.~W. Utami, H.-S. Goan, and G.~J. Milburn, Phys. Rev. B {\bf 70},  075303
  (2004).

\bibitem{goan_continuous}
H.-S. Goan, G.~J. Milburn, H.~M. Wiseman, and H.~B. Sun, Phys. Rev. B {\bf 63},
   125326  (2001).

\bibitem{goan_dynamics}
H.-S. Goan and G.~J. Milburn, Phys. Rev. B {\bf 64},  235307  (2001).

\bibitem{goan_montecarlo}
H.-S. Goan, Phys. Rev. B {\bf 72},  075305  (2005).

\bibitem{glendinning}
P. Glendinning, {\em Stability, instability and chaos: an introduction to the
  theory of nonlinear differential equations} (Cambridge University Press, New
  York, 1994).

\bibitem{carmichael_open}
H.~J. Carmichael, {\em An open systems approach to quantum optics}
  (Springer-Verlag, Berlin, Heidelberg, 1993).

\bibitem{golub}
G.~H. Golub and C.~F. Loan, {\em Matrix Computations}, 3rd ed. (The John
  Hopkins University Press, Baltimore, Maryland 21218, USA, 1996).

\bibitem{nord}
T. Nord, L. Gorelik, R. Shekhter, and M.Jonson, Phys. Rev. B {\bf 65},  165312
  (2002).

\bibitem{dum}
R. Dum, P. Zoller, and H. Ritsch, Phys. Rev. A {\bf 45},  4879  (1992).

\bibitem{kouwenhoven}
J.~M. Elzerman {\it et~al.}, Nature {\bf 403},  431  (2004).

\bibitem{sun}
H.~B. Sun and G. Milburn, Phys. Rev. B {\bf 59},  10748  (1999).

\bibitem{armour_current}
A. Armour, Phys. Rev. B {\bf 70},  165315  (2004).

\end{thebibliography}

\end{document}